%% file: vldb_sample.tex
\documentclass{vldb}
\usepackage{graphicx}
\usepackage{balance}  
\usepackage{comment}
\usepackage{epsfig}
\usepackage{epstopdf}
\usepackage{graphics}
\usepackage{hyperref}
\usepackage{amssymb}
\usepackage{amsmath}
\usepackage{booktabs} 
\usepackage{bookmark}
\usepackage{MnSymbol}%
\usepackage{multirow}

\usepackage[ruled, boxed,vlined,linesnumbered]{algorithm2e}
\usepackage{enumitem}
\usepackage{xcolor}
\usepackage{xspace}
\usepackage{algorithmic}
\usepackage{subcaption}
\usepackage{verbatim}
\usepackage{mathtools}
\usepackage{upgreek}
\usepackage{multicol,xparse,environ}
\usepackage{caption}
\usepackage{subcaption}
\usepackage[nodisplayskipstretch]{setspace}

\NewEnviron{auxmulticols}[1]{%
  \ifnum#1<2\relax
    \BODY
  \else
    \begin{multicols}{#1}
      \BODY
    \end{multicols}%
  \fi
}

\hypersetup{
  colorlinks   = true, 
  urlcolor     = black, 
  linkcolor    = red, 
  citecolor   = blue 
}

\newcommand\BbbGammaVar{\reflectbox{\rotatebox[origin=c]{180}{$\mathbb L$}}}

\vldbTitle{Privacy Preserving Vertical Federated Learning for Tree-based Models}
\vldbAuthors{Yuncheng Wu, Shaofeng Cai, Xiaokui Xiao, Gang Chen, Beng Chin Ooi}
\vldbDOI{https://doi.org/10.14778/3407790.3407811}
\vldbVolume{13}
\vldbNumber{11}
\vldbYear{2020}

\begin{document}


\title{Privacy Preserving Vertical Federated Learning for Tree-based Models \\
\Large [Technical Report]}

\numberofauthors{5}
\author{
{Yuncheng Wu$^\dag$, Shaofeng Cai$^\dag$, Xiaokui Xiao$^\dag$, Gang Chen$^\ddag$, Beng Chin Ooi$^\dag$}\\
\fontsize{11}{11}\selectfont\ttfamily \hspace{0mm}
$^\dag$National University of Singapore  $^\ddag$Zhejiang University \hspace{18mm}\\
\fontsize{11}{11}\selectfont\ttfamily\upshape\vspace{-0.0em}\hspace{10mm}
\{wuyc, shaofeng, xiaoxk, ooibc\}@comp.nus.edu.sg   cg@zju.edu.cn \hspace{12mm}
}


\maketitle

\begin{abstract}


\par

Federated learning (FL) is an emerging paradigm that enables multiple organizations to jointly train a model without revealing their private data to each other. This paper studies {\it vertical} federated learning, which tackles the scenarios where (i) collaborating organizations own data of the same set of users but with disjoint features,  and (ii) only one organization holds the labels. We propose Pivot, a novel solution for privacy preserving vertical decision tree training and prediction, ensuring that no intermediate information is disclosed other than those the clients have agreed to release (i.e., the final tree model and the prediction output).
Pivot does not rely on any trusted third party and provides protection against a semi-honest adversary that may compromise $m-1$ out of $m$ clients. We further identify two privacy leakages when the trained decision tree model is released in plaintext and propose an enhanced protocol to mitigate them.
The proposed solution can also be extended to tree ensemble models, e.g., random forest (RF) and gradient boosting decision tree (GBDT) by treating single decision trees as building blocks.
Theoretical and experimental analysis suggest that Pivot is efficient for the privacy achieved.

\end{abstract}

\input{tex-files/sec-introduction.tex}
\input{tex-files/sec-background.tex}
\input{tex-files/sec-system-overview.tex}

\input{tex-files/sec-decision-trees.tex}
\input{tex-files/sec-enhanced-solution.tex}
\input{tex-files/sec-analysis.tex}
\input{tex-files/sec-extensions.tex}
\input{tex-files/sec-experiments.tex}
\input{tex-files/sec-further-extension.tex}
\input{tex-files/sec-related-work.tex}

\input{tex-files/sec-conclusion.tex}
\input{tex-files/sec-acknowledgements.tex}

\balance

\bibliographystyle{abbrv}
\bibliography{vldb_sample.bib}  



\end{document}

%% file: tex-files/sec-introduction.tex
\section{Introduction}\label{sec:introduction}

There has been a growing interest in exploiting data from distributed databases of multiple organizations, for providing better customer service and acquisition.
\textit{Federated learning} (FL) \cite{McMahanMRA16,McMahanRT018} (or \textit{collaborative learning} \cite{HitajAP17}) is an emerging paradigm for machine learning that enables multiple data owners (i.e., \textit{clients}) to jointly train a model without revealing their private data to each other. The basic idea of FL is to iteratively let each client (i) perform some local computations on her data to derive certain intermediate results, and then (ii) exchange these results with other clients in a secure manner to advance the training process, until a final model is obtained. The advantage of FL is that it helps each client protect her data assets, so as to abide by privacy regulations (e.g., GDPR \cite{GDPR2016} and CCPA \cite{CCPA2018}) or to maintain a competitive advantage from proprietary data.

\begin{figure}[t]
    \centering
    \includegraphics[width=0.96\columnwidth]{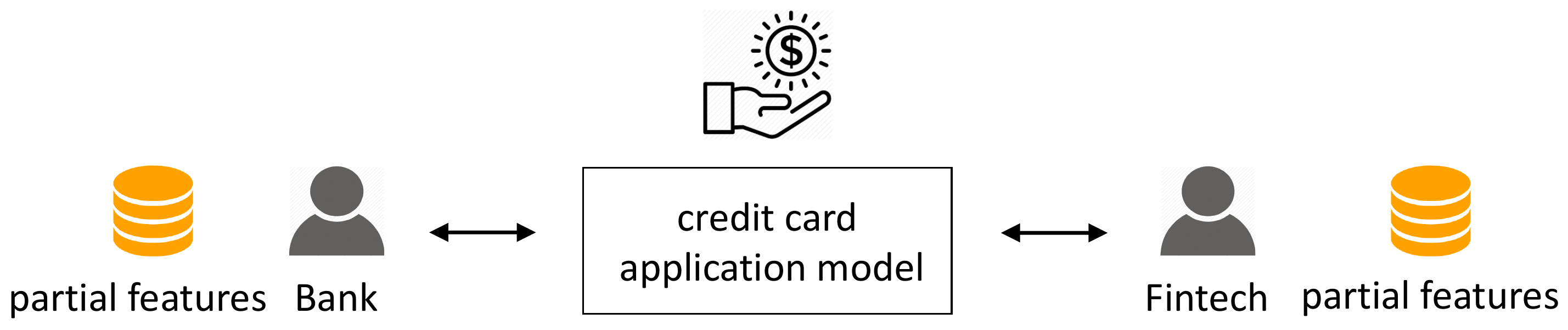}
    \caption{Example of vertical federated learning}
    \label{fig:vfl-example}
\end{figure}

Existing work on FL has mainly focused on the {\it horizontal} setting \cite{Bonawitz17, McMahanMRA16, McMahanRT018, Zheng2019, Shokri15, OhrimenkoSFMNVC16, BhagojiCMC19, NikolaenkoWIJBT13, MelisSCS19}, which assumes that each client's data have the same schema, but no tuple is shared by multiple clients. In practice, however, there is often a need for {\it vertical federated learning}, where all clients hold the same set of records, while each client only has a disjoint subset of features.
For example, Figure~\ref{fig:vfl-example} illustrates a digital banking scenario, where a bank and a Fintech company aim to jointly build a machine learning model that evaluates credit card applications.
The bank has some partial information about the users (e.g., account balances), while the Fintech company has some other information (e.g., the users' online transactions).
In this scenario, vertical FL could enable the bank to derive a more accurate model, while the Fintech company could benefit from a pay-per-use model \cite{WuZPM16} for its contribution to the training and prediction.

To our knowledge, there exist only a few  solutions \cite{WangXSY06, HuNYZ19, VaidyaC05, VaidyaCKP08, VaidyaSFML14, ChengCorr19, LiuLL19, OhrimenkoSFMNVC16} for privacy preserving vertical FL. These solutions, however, are insufficient in terms of either efficiency or data privacy.  
In particular, \cite{WangXSY06, HuNYZ19} assume that the labels in the training data could be shared with all participating clients in plaintext, whereas in practice, the labels often exist in one client's data only and could not be revealed to other clients without violating privacy. For instance, in the scenario illustrated in Figure~\ref{fig:vfl-example}, the training data could be a set of historical credit card applications, and each label would be a ground truth that indicates whether the application should have been approved. In this case, the labels are only available to the bank and could not be directly shared with the Fintech company. As a consequence, the solutions in \cite{WangXSY06, HuNYZ19} are inapplicable.
Meanwhile, \cite{VaidyaC05, VaidyaCKP08, VaidyaSFML14, ChengCorr19, LiuLL19} assume that some intermediate results during the execution could be revealed in plaintext; nevertheless,
such intermediate results could be exploited by an adversarial client to infer the sensitive information in other clients' data.
The solution in \cite{OhrimenkoSFMNVC16}, on the other hand, relies on secure hardware \cite{McKeenABRSSS13} for privacy protection, but such secure hardware may not be trusted by all parties \cite{Zheng2019} and could be vulnerable to side channel attacks \cite{XuCP15}.
The method in \cite{Mohassel17} utilizes secure multiparty computation (MPC) \cite{Yao82b}, but assumes that each client's data could be outsourced to a number of non-colluding servers. This assumption is rather strong, as it is often challenging in practice to ensure that those servers do not collude and to convince all clients about it.

To address the above issues, we propose {\it Pivot}, a novel and efficient solution for vertical FL that does not rely on any trusted third party and provides protection against a semi-honest adversary that may compromise $m-1$ out of $m$ clients.
Pivot is a part of our {\it Falcon}\footnote{\url{https://www.comp.nus.edu.sg/~dbsystem/fintech/project/falcon/}} (federated learning with privacy protection) system, and it ensures that no intermediate information is disclosed during the training or prediction process.
Specifically, Pivot is designed for training decision tree (DT) models, which are well adopted for financial risk management \cite{ChengCorr19, LiuLL19}, healthcare analytics \cite{AzarE13a}, and fraud detection \cite{CaoYCZLQ19} due to their good interpretability. 
The core of Pivot is a hybrid framework that utilizes both threshold partially homomorphic encryption (TPHE) and MPC, which are two cryptographic techniques that complement each other especially in the vertical FL setting: TPHE is relatively efficient in terms of communication cost but can only support a restrictive set of computations, whereas MPC could support an arbitrary computation but incurs expensive communication overheads.
Pivot employs TPHE as much as possible to facilitate clients' local computation, and only invokes MPC in places where TPHE is inadequate in terms of functionality. This leads to a solution that is not only secure but also highly efficient for vertical tree models, as demonstrated in Section~\ref{sec:experiments}.
Specifically, we make the following contributions:
\begin{itemize}[topsep=2pt,itemsep=2pt,parsep=0pt,partopsep=0pt,leftmargin=15pt]

\item We propose a basic protocol of Pivot that supports the training of both classification trees and regression trees, as well as distributed prediction using the tree models obtained. This basic protocol guarantees that each client only learns the final tree model but nothing else. To our knowledge, Pivot is the first vertical FL solution that achieves such a guarantee.

\item We enhance the basic protocol of Pivot to handle a more stringent case where parts of the final tree model need to be concealed for better privacy protection. In addition, we propose extensions of Pivot for training several {\it ensemble} tree-based models, including random forest (RF) and gradient boosting decision trees (GBDT).

\item We implement DT, RF, and GBDT models based on Pivot and conduct extensive evaluations on both real and synthetic datasets. The results demonstrate that Pivot offers accuracy comparable to non-private algorithms and provides high efficiency. The basic and enhanced protocols of Pivot achieve up to 37.5x and 4.5x speedup (\textit{w.r.t.} training time) over an MPC baseline.

\end{itemize}

%% file: tex-files/sec-background.tex
\vspace{-0.3cm}
\section{Preliminaries}\label{sec:preliminaries}
\vspace{-1mm}
\subsection{Partially Homomorphic Encryption}
\label{subsec:preliminaries:threshold-homomorphic-encryption}
A partially homomorphic encryption (PHE) scheme is a probabilistic asymmetric encryption scheme for restricted computation over the ciphertexts. 
In this paper, we utilize the Paillier cryptosystem \cite{Paillier99}, which consists of three algorithms (\textbf{Gen, Enc, Dec}):
\begin{itemize}[topsep=2pt,itemsep=2pt,parsep=0pt,partopsep=0pt,leftmargin=15pt]
\item The \textit{key generation} algorithm $(sk, pk)=\textbf{Gen}(keysize)$ which returns secret key $sk$ and public key $pk$, given a security parameter \textit{keysize}. 
\item The \textit{encryption} algorithm $c = \textbf{Enc}(x, pk)$, which maps a plaintext $x$ to a ciphertext $c$ using $pk$.
\item The \textit{decryption} algorithm $x = \textbf{Dec}(c, sk)$, which reverses the encryption by $sk$ and outputs the plaintext $x$. 
\end{itemize}
Interested readers are referred to \cite{Damg01} for the exact construction of \textbf{Enc} and \textbf{Dec}. For simplicity, we omit the public key $pk$ in the \textbf{Enc} algorithm and write $\textbf{Enc}(x)$ as $[x]$ in the rest of the paper. Let $x_1, x_2$ denote two plaintexts. We utilize the following properties of PHE:

\vspace{1mm}
\noindent 
\textbf{Homomorphic addition:} given two ciphertexts $[x_1]$, $[x_2]$, the ciphertext of the sum $x_1+x_2$ can be obtained by multiplying the ciphertexts, i.e.,
\begin{align}
\label{eq:homomorphic-addition}
[x_1] \oplus [x_2]:\  [x_1] \cdot [x_2] = [x_1+x_2]
\end{align} \par

\vspace{0mm}
\noindent
\textbf{Homomorphic multiplication:} given a plaintext $x_1$ and a ciphertext $[x_2]$,
the ciphertext of the product $x_1x_2$ can be obtained by raising $[x_2]$ to the power $x_1$:
\begin{align}
x_1 \otimes [x_2] : \ [x_2]^{x_1} = [x_1 x_2]
\end{align} \par

\vspace{0mm}
\noindent
\textbf{Homomorphic dot product:} given a ciphertext vector $[\boldsymbol{v}]=([v_1],\cdots,[v_m])^T$ and a plaintext vector $\boldsymbol{x}=(x_1,\cdots,x_m)$, the ciphertext of the dot product $\boldsymbol{v} \cdot \boldsymbol{x}$ can be obtained by:
\begin{align}\label{eq:dot-product}
\boldsymbol{x} \odot [\boldsymbol{v}] \ & :  (x_1 \otimes [v_1]) \oplus \cdots \oplus (x_m \otimes [v_m])  \nonumber \\
& = [x_1 v_1  + \cdots + x_m v_m] \\
& = [\boldsymbol{x} \cdot \boldsymbol{v}] \nonumber
\end{align}

We utilize a {\it threshold variant} of the PHE scheme (i.e., TPHE) with the following additional properties. First, the public key $pk$ is known to everyone, while each client only holds a partial secret key. 
Second, the decryption of a ciphertext requires inputs from a certain number of clients. 
In this paper, we use a {\it full threshold structure}, which requires all clients to participate in order to decrypt a ciphertext. 

\vspace{-2mm}
\subsection{Secure Multiparty Computation}\label{subsec:preliminaries:secure-multiparty-computation}

Secure multiparty computation (MPC) allows participants to compute a function over their inputs while keeping the inputs private. In this paper, we utilize the additive secret sharing scheme SPDZ \cite{DamgardPSZ12} for MPC. 
We refer to a value $a \in \mathbb{Z}_q$ that is additively shared among clients as a \textit{secretly shared value}, and denote it as $\langle a \rangle = ({\langle a \rangle}_1, \cdots, {\langle a \rangle}_m)$, where ${\langle a \rangle}_i$ is a random \textit{share} of $\langle a \rangle$ hold by client $i$. 
To reconstruct a secretly shared value $\langle a \rangle$, i.e., $\textbf{Rec}(\langle a \rangle)$, every client can send its own share to a specific client who computes $a = (\sum_{i = 1}^{m} {\langle a \rangle}_i) \mod q$. Given secretly shared values, we have the following secure computation primitives. For ease of exposition, we omit the modular operation in the following formulations.

\vspace{1mm}
\noindent
\textbf{Secure addition:} given two secretly shared values $\langle a \rangle$ and $\langle b \rangle$, the secretly shared sum $c = a + b$ can be obtained by having client $i$ non-interactively compute ${\langle c \rangle}_i = {\langle a \rangle}_i + {\langle b \rangle}_i$. Then ${\langle c \rangle}_i$ is a share of $\langle c \rangle$ owned by client $i$. \par 

\vspace{1mm}
\noindent
\textbf{Secure multiplication:} given two secretly shared values $\langle a \rangle$ and $\langle b \rangle$, the secretly shared multiplication $c = a \cdot b$ can be obtained using Beaver's pre-computed multiplication triplet technique \cite{Beaver91a}. Assuming that the clients have already shared $\langle u \rangle$, $\langle v \rangle$, $\langle z \rangle$ where $u, v$ are random values in $\mathbb{Z}_q$ and $z = u \cdot v \mod q$, then client $i$ locally computes ${\langle e \rangle}_i = {\langle a \rangle}_i - {\langle u \rangle}_i$ and ${\langle f \rangle}_i = {\langle b \rangle}_i - {\langle v \rangle}_i$, and the clients run $\textbf{Rec}(\langle e \rangle)$ and $\textbf{Rec}(\langle f \rangle)$. Finally, every client $i$ computes ${\langle c \rangle}_i = -i \cdot e \cdot f + f \cdot {\langle a \rangle}_i + e \cdot {\langle b \rangle}_i + {\langle z \rangle}_i$. Then ${\langle c \rangle}_i$ is a share of $\langle c \rangle$ owned by client $i$. \par  

\vspace{1mm}
\noindent
\textbf{Secure comparison:} given two shared values $\langle a \rangle$ and $\langle b \rangle$, the secure comparison operation (e.g., $\langle a \rangle > \langle b \rangle$) returns a secretly shared $\langle 0 \rangle$ or $\langle 1 \rangle$. The basic idea is to first truncate the two values by $2^k$, then execute $\langle a \rangle - \langle b \rangle$, and finally output the secretly shared sign bit after dividing by $2^{k-1}$. We refer the interested readers to \cite{CatrinaH10,CatrinaS10} for details.

{Based on the above primitives, other primitives including secure division and secure exponential can be approximated, which are also supported in SPDZ \cite{CatrinaS10, DamgardPSZ12, ArakiB0KLOT18}.
In this paper, we use these secure computations in SPDZ as building blocks by default for the calculation concerning secretly shared values, which means that the outputs are also secretly shared values unless they are reconstructed.}
The secret sharing based MPC has two phases: an offline phase that is independent of the function and generates pre-computed Beaver's triplets, and an online phase that computes the designated function using these triplets.

\IncMargin{0.5em}
\begin{algorithm}[t]
\DontPrintSemicolon
\small
\KwIn{${F}$: feature set,
${Y}$: label set, 
${D}$: sample set
}
\KwOut{$T$: decision tree}
{
    \If{prune conditions satisfied} {
        classification: return leaf node with majority class \\
        regression: return leaf node with mean label value
    }
    \Else {
        determine the best split feature $j$ and value $s$ \\
        split ${D}$ into 2 partitions ${D}_l, {D}_r$ \\
        return a tree with feature $a$ that has two edges, call CART($F-j,Y, D_l$) and CART($F-j,Y,D_r$)
    }
}
\caption{CART($F,Y,D$)}\label{alg:cart}
\end{algorithm}

\vspace{-2mm}
\subsection{Tree-based Models}\label{subsec:preliminaries:tree-models}

In this paper, we consider the classification and regression trees (CART) algorithm \cite{BreFriOlsSto84a} with binary structure, while we note that other variants (e.g., ID3 \cite{Quinlan1986}, C4.5 \cite{Quinlan1993}) can be easily generalized. We assume there is a training dataset ${D}$ with $n$ data points $\{\textbf{x}_1, \cdots, \textbf{x}_n \}$ each containing $d$ features and the corresponding output label set $Y = \{y_1, \cdots, y_n \}$. 

Algorithm \ref{alg:cart} describes the CART algorithm, which builds a tree recursively.
For each tree node, it first decides whether some pruning conditions are satisfied, e.g., feature set is empty, tree reaches the maximum depth, the number of samples is less than a threshold.
If any condition is satisfied, then it returns a leaf node with the class of majority samples for classification or the mean label value for regression.
Otherwise, it determines the best split to construct two sub-trees that are built recursively.
In order to find the best split feature and split threshold, CART uses Gini impurity \cite{BreFriOlsSto84a} as a metric in classification. Let $c$ be the number of classes and $K = \{1,\cdots,c\}$ be the class set. Let $D$ be sample set on a given node, the Gini impurity is:
\begin{align}
    \label{eq:gini-impurity}
    I_G (D) = 1 - \sum\nolimits_{k\in K} (p_k)^2
\end{align}
where $p_k$ is the fraction of samples in $D$ labeled with class $k$. Let $F$ be the set of available features, given any split feature $j \in F$ and split value $\tau \in \text{Domain}(j)$, the sample set $D$ can be split into two partitions $D_l$ and $D_r$. Then, the {\it impurity gain} of this split is as follows:
\begin{align}
    \label{eq:best-split}
    \text{\it gain} & = I_G(D) - \left(w_l \cdot I_G(D_l) + w_r \cdot I_G(D_r)\right) \nonumber \\
    & = w_l \sum\nolimits_{k \in K} (p_{l,k})^2 + w_r \sum\nolimits_{k \in K} (p_{r,k})^2 - \sum\nolimits_{k \in K} (p_k)^2
\end{align}
where $w_l = |D_l|/|D|$ and $w_r = |D_r|/|D|$, and $p_{l,k}$ (resp.\ $p_{r,k}$) is the fraction of samples in $D_l$ (resp.\ $D_r$) that are labeled with class $k \in K$. The split with the maximum impurity gain is considered the best split of the node. 
For regression, CART uses the label variance as a metric. Let $Y$ be the set of labels of $D$, then the label variance is:
\begin{align}
    \label{eq:variance}
    I_V(D) = E(Y^2) - (E(Y))^2 = \frac{1}{n} \sum_{i=1}^n y_i^2 - (\frac{1}{n} \sum_{i=1}^n y_i)^2 
\end{align}
Similar to Eqn (\ref{eq:best-split}), the best split is determined by maximizing the variance gain. With CART, ensemble models can be trained to obtain better predictive performance, such as random forest (RF) \cite{Breiman01}, gradient boosting decision tree (GBDT) \cite{Friedman00, FuJSC19}, XGBoost \cite{ChenG16}, etc.

%% file: tex-files/sec-system-overview.tex
\section{Solution Overview}\label{sec:system-overview}

\subsection{System Model}\label{subsec:system-model}

We consider a set of $m$ distributed clients (or data owners) $\{u_1, \cdots, u_m\}$ who want to train a decision tree model by consolidating their respective dataset $\{{D}_1, \cdots, {D}_m\}$. Each row in the datasets corresponds to a data sample, and each column corresponds to a feature. Let $n$ be the number of samples and $d_i$ be the number of features in ${D}_i$, where $i \in \{1, \cdots, m\}$. We denote $D_i = \{\textbf{x}_{it}\}_{t=1}^{n}$ where $\textbf{x}_{it}$ represents the $t$-th sample of $D_i$. Let $Y = \{y_t\}_{t=1}^n$ be the set of sample labels. 
Table \ref{table:notations} summarizes the frequently used notations. \par

Pivot focuses on the vertical federated learning scenario \cite{YangLCT19}, where the datasets $\{{D}_1, \cdots, D_m\}$ share the same sample ids while with different features. 
In particular, we assume that the clients have determined and aligned their common samples using private set intersection techniques \cite{Meadows86, Pinkas0SZ15, ChenLR17, Pinkas18} without revealing any information about samples not in the intersection. 
In addition, we assume that the label set $Y$ is held by only one client (i.e., \textit{super client}) and cannot be directly shared with other clients.

\subsection{Threat Model}\label{subsec:threat-model}

We consider the semi-honest model \cite{Mohassel17, WuWZLCL18, Wu2019, CockDHKNPT19, Chowdhury19, 0002IK19} where every client follows the protocol exactly as specified, but may try to infer other clients' private information based on the messages received. 
Like any other client, no additional trust is assumed of the super client.
We assume that an adversary $\mathcal{A}$ can corrupt up to $m-1$ clients
and the adversary's corruption strategy is static, such that the set of corrupted clients is fixed before the protocol execution and remains unchanged during the execution.

\subsection{Problem Formulation}\label{subsec:problem-formulation}

To protect the private data of honest clients, we require that an adversary learns nothing more than the data of the clients he has corrupted and the final output.
Similar to previous work \cite{Mohassel17, Zheng2019, CockDHKNPT19}, we formalize our problem under 
the ideal/real paradigm. Let $\mathcal{F}$ be an ideal functionality such that the clients send their data to a trusted third party for computation and receive the final output from that party. Let $\pi$ be a real world protocol executed by the clients. 
We say a real protocol $\pi$ behaviors indistinguishably as the ideal functionality $\mathcal{F}$ if the following formal definition is satisfied.

\newdef{definition}{Definition}
\begin{definition}\label{def:uc-model}
(\cite{Canetti01,Cramer2015,Mohassel17}). A protocol $\pi$ securely realizes an ideal functionality $\mathcal{F}$ if for every adversary $\mathcal{A}$ attacking the real interaction, there exists a simulator $\mathcal{S}$ attacking the ideal interaction, such that for all environments $\mathcal{Z}$, the following quantity is negligible (in $\lambda$):
\begin{align*}
    \big|\text{Pr}[\textsc{real}(\mathcal{Z}, \mathcal{A}, \pi, \lambda) = 1] - \text{Pr}[\textsc{ideal}(\mathcal{Z}, \mathcal{S}, \mathcal{F}, \lambda) = 1] \big|. \Box
\end{align*}
\end{definition}
In this paper, we identify two ideal functionalities $\mathcal{F}_{\text{DTT}}$ and $\mathcal{F}_{\text{DTP}}$ for the model training and model prediction, respectively. In $\mathcal{F}_{\text{DTT}}$, the input is every client's dataset while the output is the trained model that all clients have agreed to release. In $\mathcal{F}_{\text{DTP}}$, the input is the released model and a sample while the output is the predicted label of that sample. The output of $\mathcal{F}_{\text{DTT}}$ is part of the input of $\mathcal{F}_{\text{DTP}}$. 
Specifically, in our basic protocol (Section~\ref{sec:decision-trees}), we assume that the output of $\mathcal{F}_{\text{DTT}}$ is the plaintext tree model, including the split feature and the split threshold on each internal node, and the label for prediction on each leaf node. 
While in our enhanced protocol (Section~\ref{sec:enhanced-solution}), the released plaintext information is assumed to include only the split feature on each internal node, whereas the split threshold and the leaf label are concealed for better privacy protection.

\begin{table}[t]
\caption{Summary of notations}
\vspace{-0.2cm}
\small
\centering
\begin{tabular}{  l  l }
\toprule
Notation & Description \\
\toprule
$m$ & number of clients \\
$n$ & number of samples in the training dataset \\
$d$ & number of total features \\
${D}_i$ & training dataset hold by client $i$ \\
$Y$ & label set of training dataset \\
$d_i$ & number of features in ${D}_i$ \\
$b$ & maximum split number for any feature \\
$h$ & maximum tree depth \\
$pk, sk$ & public key and secret key pair \\
$[\boldsymbol{\alpha}]$ & encrypted mask vector for a tree node \\
\bottomrule
\end{tabular}
\label{table:notations}
\end{table}

\subsection{Protocol Overview}\label{subsec:Solution-overview}

We now provide the protocol overview of Pivot. The protocols are composed of three stages: initialization, model training, and model prediction. \par

\vspace{1mm}
\noindent
\textbf{Initialization stage.} In this stage, the $m$ clients agree to run a designated algorithm (i.e., the decision tree model) over their joint data and release the pre-defined information (e.g., the trained model) among themselves. The clients collaboratively determine and align the joint samples.
The clients also build consensus on some hyper-parameters, such as security parameters (e.g., key size), pruning thresholds, and so on. The $m$ clients jointly generate the keys of threshold homomorphic encryption and every client $u_i$ receives the public key $pk$ and a partial secret key ${sk}_i$. 

\vspace{1mm}
\noindent
\textbf{Model training stage.} The $m$ clients build the designated tree model iteratively. In each iteration, the super client first broadcasts some encrypted information to facilitate the other clients to compute encrypted necessary statistics at local.
After that, the clients jointly convert those statistics into MPC-compatible inputs, i.e., secretly shared values, to determine the best split of the current tree node using secure computations. 
Finally, the secretly shared best split is revealed (in the Pivot basic protocol) or is converted back into an encrypted form (in the Pivot enhanced protocol), for clients to update the model. 
Throughout the whole process, no intermediate information is disclosed to any client. 

\vspace{1mm}
\noindent
\textbf{Model prediction stage.} After model training, the clients obtain a tree model. In the basic protocol of Pivot (Section~\ref{sec:decision-trees}), the whole tree is released in plaintext. In the Pivot enhanced protocol (Section~\ref{sec:enhanced-solution}), the split threshold on each internal node and the prediction label on each leaf node are concealed from all clients, in secretly shared form. Given an input sample with distributed feature values, the clients can jointly produce a prediction. Pivot guarantees that no information except for the predicted label is revealed during the prediction process.

%% file: tex-files/sec-decision-trees.tex
\section{Basic Protocol}\label{sec:decision-trees}

In this section, we present our basic protocol of Pivot. The output of the model training stage is assumed to be the whole plaintext tree model.
Note that prior work \cite{WangXSY06, HuNYZ19, VaidyaC05, VaidyaCKP08, VaidyaSFML14, ChengCorr19, LiuLL19} is not applicable to our problem since they simplify the problem by revealing either the training labels or intermediate results in plaintext, which discloses too much information regarding the client's private data.

To satisfy Definition~\ref{def:uc-model} for vertical tree training, a straightforward solution is to directly use the MPC framework. For example, the clients can apply the additive secret sharing scheme (see Section~\ref{subsec:preliminaries:secure-multiparty-computation}) to convert private datasets and labels into secretly shared data, and train the model by secure computations.
However, this solution incurs high communication complexity because it involves $O(nd)$ secretly shared values and most secure computations are communication intensive.
On the other hand, while TPHE could enable each client to compute encrypted split statistics at local by providing the super client's encrypted label information, it does not support some operations (e.g., comparison), which are needed in best split determination.
Based on these observations and inspired by \cite{Zheng2019}, we design our basic protocol using a hybrid framework of TPHE and MPC for vertical tree training. The basic idea is that each client executes as many local computations (e.g., computing split statistics) as possible with the help of TPHE and uses MPC only when TPHE is insufficient (e.g., deciding the best split).
As a consequence, most computations are executed at local and the secretly shared values involved in MPC are reduced to $O(d b)$, where $b$ denotes the maximum number of split values for any feature and $d b$ is the number of total splits.

Section~\ref{subsec:classification-train} and Section~\ref{subsec:regression-train} present our training protocol for classification tree and regression tree, respectively. Section~\ref{subsec:prediction-trees} proposes our tree model prediction method. The security analysis is provided in Section~\ref{subsec:privacy-analysis}.

\subsection{Classification Tree Training}\label{subsec:classification-train}

In our training protocol, the clients use an \textit{mask vector} of size $n$ to indicate which samples are available on a tree node, but keep the vector in an encrypted form to avoid disclosing the sample set. Specifically, let $\boldsymbol{\alpha} = (\alpha_1, \cdots, \alpha_n)$ be an indicator vector for a tree node. Then, for any $i \in \{1,\cdots,n\}$, $\alpha_i = 1$ indicates that the $i$-th sample is available on the node, and $\alpha_i = 0$ otherwise. We use $[\boldsymbol{\alpha}] = ([\alpha_1], \cdots, [\alpha_n])$ to denote the encrypted version of $\boldsymbol{\alpha}$, where $[\cdot]$ represents homomorphic encrypted values (see Section~\ref{subsec:preliminaries:threshold-homomorphic-encryption}).

Before the training starts, each client initializes a decision tree with only a root node, and associates the root node with an encrypted indicator vector $[\boldsymbol{\alpha}]$ where all elements are $[1]$ (since all samples are available on the root node). Then, the clients work together to recursively split the root node.
In what follows, we will use an example to illustrate how our protocol decides the best split for a given tree node based on Gini impurity.

\begin{figure}[t]
\centering
\includegraphics[width=0.96\columnwidth]{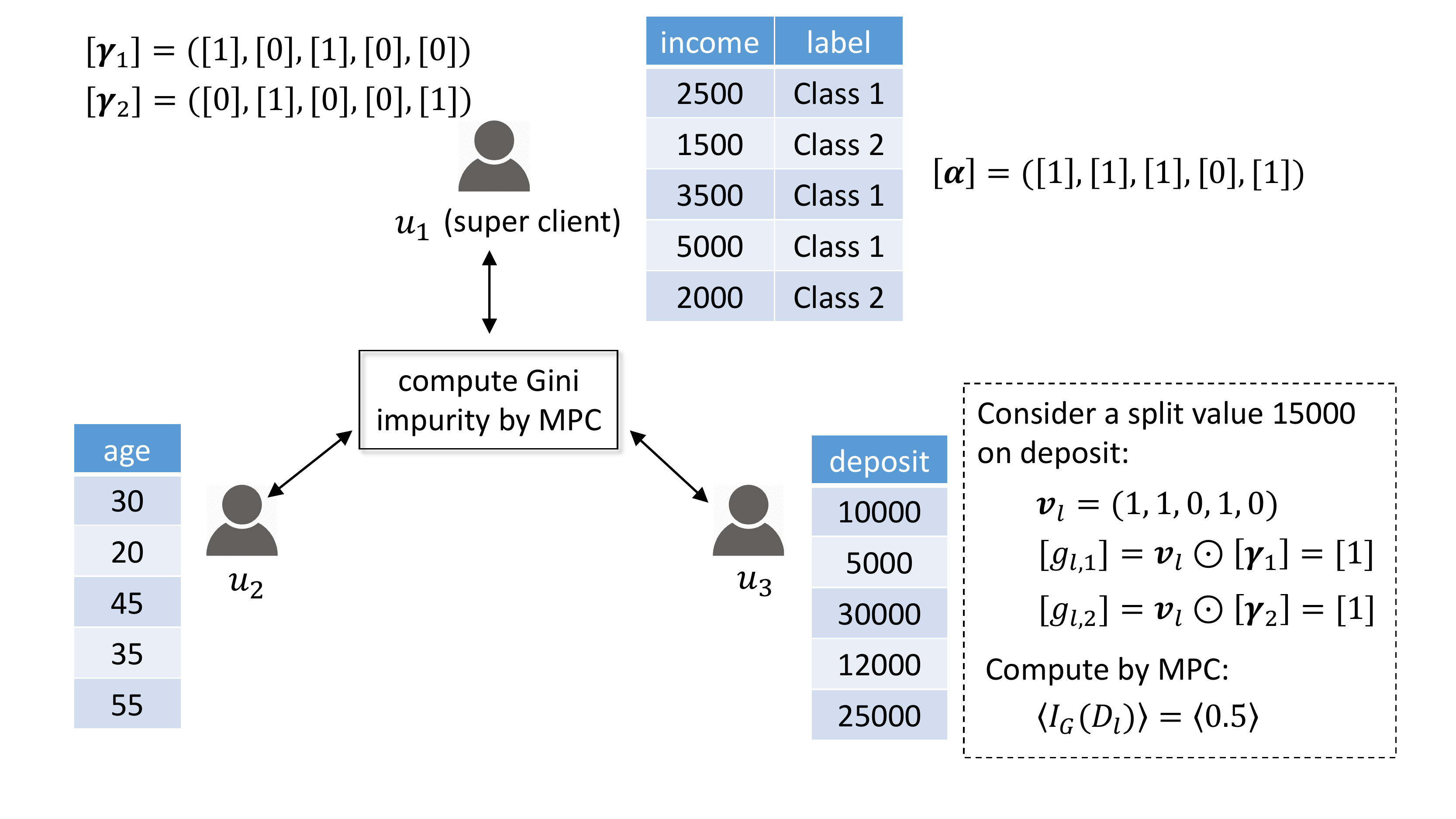}
\vspace{-2mm}
\caption{Classification tree training example}
\label{fig:classification-example}
\end{figure}

Consider the example in Figure~\ref{fig:classification-example}, where we have three clients $u_1$, $u_2$, and $u_3$. Among them, $u_1$ is the super client, and she owns the labels with two classes, $1$ and $2$. There are five training samples with three features (i.e., income, age, and deposit), and each client holds one feature. 
Suppose that the clients are to split a tree node whose encrypted mask vector is $[\boldsymbol{\alpha}] = ([1],[1],[1],[0],[1])$, i.e., Samples 1, 2, 3, and 5 are on the node.
Then, $u_1$ computes an encrypted indicator vector for each class, based on $[\boldsymbol{\alpha}]$ and her local labels. For example, for Class 1, $u_1$ derives an temporary indicator vector $(1,0,1,1,0)$, which indicates that Samples 1, 3, and 4 belong to Class 1. Next, $u_1$ uses the indicator vector to perform an element-wise homomorphic multiplication with $[\boldsymbol{\alpha}]$, which results in an encrypted indicator vector $[\boldsymbol{\gamma}_1] = ([1],[0],[1],[0],[0])$.
This vector indicates that Samples 1 and 3 are on the node to be split, and they belong to Class 1. Similarly, $u_1$ also generates an encrypted indicator vector $[\boldsymbol{\gamma}_2]$ for Class 2. After that, $u_1$ broadcasts $[\boldsymbol{\gamma}_1]$ and $[\boldsymbol{\gamma}_2]$ to all clients. 

After receiving $[\boldsymbol{\gamma}_1]$ and $[\boldsymbol{\gamma}_2]$, each client combines them with her local training data to compute several statistics that are required to choose the best split of the current node. In particular, to evaluate the quality of a split based on Gini impurity (see Section~\ref{subsec:preliminaries:tree-models}), each client needs to examine the two child nodes that would result from the split, and then compute the following statistics for each child node: (i) the total number of samples that belong to the child node, and (ii) the number of samples among them that are associated with label class $k$, for each $k \in K$.

For example, suppose that $u_3$ considers a split that divides the current node based on whether the deposit values are larger than 15000. Then, $u_3$ first examines her local samples, and divide them into two partitions. The first partition (referred to as the {\it left partition}) consists of Samples 1, 2, and 4, i.e., the local samples whose deposit values are no more than 15000. Meanwhile, the second partition (referred to as the {\it right partition}) contains Samples 3 and 5. Accordingly, for the left (resp.\ right) partition, $u_3$ constructs an indicator vector $\boldsymbol{v}_l = (1, 1, 0, 1, 0)$ (resp.\ $\boldsymbol{v}_r = (0, 0, 1, 0, 1)$) to specify the samples that it contains. 
After that, $u_3$ performs a homomorphic dot product between $\boldsymbol{v}_l$ and $[\boldsymbol{\gamma}_1]$ to obtain an encrypted number $[g_{l, 1}]$. 
Observe that $g_{l, 1}$ equals the exact number of Class 1 samples that belong to the left child node of the split. Similarly, $u_3$ uses $\boldsymbol{v}_l$ and $[\boldsymbol{\gamma}_2]$ to generate $[g_{l,2}]$, an encrypted version of the number of Class 2 samples that belong to the left child node. Using the same approach, $u_3$ also computes the encrypted numbers of Classes 1 and 2 samples associated with the right child node. Further, $u_3$ derives an encrypted total number of samples in the left (resp.\ right) child node, using a homomorphic dot product between $\boldsymbol{v}_l$ and $[\boldsymbol{\alpha}]$  (resp.\ $\boldsymbol{v}_r$ and $[\boldsymbol{\alpha}]$).

Suppose that each client computes the encrypted numbers associated with each possible split of the current node, using the approach illustrated for $u_3$ above. Then, they can convert them into MPC-compatible inputs, and then invoke an MPC protocol to securely identify the best split of the current node. We will elaborate the details shortly.
 
In general, the clients split each node in three steps: local computation, MPC computation, and model update.
In the following, we discuss the details of each step. 

\vspace{1mm}
\noindent
\textbf{Local computation.} Suppose that the clients are to split a node that is associated with an encrypted mask vector $[\boldsymbol{\alpha}] = ([\alpha_1], \cdots, [\alpha_n])$, indicating the available samples on the node.
First, the super client constructs, for each class label $k \in K$, an auxiliary indicator vector $\boldsymbol{\beta}_k = (\beta_{k,1}, \cdots, \beta_{k,n})$, such that $\beta_{k, t} = 1$ if Sample $t$'s label is $k$, and $\beta_{k, t} = 0$ otherwise.
After that, the super client performs an element-wise homomorphic multiplication between $\boldsymbol{\beta}_k$ and $[\boldsymbol{\alpha}]$, obtaining an encrypted indicator vector $[\boldsymbol{\gamma}_k]$. Then, the super client broadcasts $[\BbbGammaVar] = \bigcup_{k\in K}\{[\boldsymbol{\gamma}_k]\}$ to the other clients. 

Upon receiving $[\BbbGammaVar]$, each client $u_i$ uses it along with her local data to derive several statistics for identifying the tree node's best split, as previously illustrated in our example. 
In particular, let $F_i$ be the set of features that $u_i$ has, and $S_{ij}$ be the set of split values for a feature $j \in F_{i}$. Then, for any split value $\tau \in S_{ij}$, $u_i$ first constructs two size-$n$ indicator vectors $\boldsymbol{v}_{l}$ and $\boldsymbol{v}_{r}$, such that (i) the $t$-th element in $\boldsymbol{v}_{l}$ equals 1 if Sample $t$'s feature $j$ is no more than $\tau$, and 0 otherwise, and (ii) $\boldsymbol{v}_{r}$ complements $\boldsymbol{v}_{l}$. 
Consider the two possible child nodes induced by the split value $\tau$. For each class $k \in K$, let $g_{l,k}$ (resp.\ $g_{r,k}$) be the number of samples labeled with class $k$ that belong to the left (resp.\ right) child node. $u_i$ computes the encrypted versions of $g_{l,k}$ and $g_{r,k}$ using homomorphic dot products (see Section~\ref{subsec:preliminaries:threshold-homomorphic-encryption}) as follows:
\begin{align}
    [g_{l,k}] = \boldsymbol{v}_{l} \odot [\boldsymbol{\gamma}_k], \quad
    [g_{r,k}] = \boldsymbol{v}_{r} \odot [\boldsymbol{\gamma}_k].
    \label{eq:statistics}
\end{align}
Let $n_l$ (resp.\ $n_r$) be the number of samples in the left (resp.\ right) child node. $u_i$ computes $[n_l] = \boldsymbol{v}_l \odot [\boldsymbol{\alpha}]$ and $[n_r] = \boldsymbol{v}_r \odot [\boldsymbol{\alpha}]$. In total, for each split value $\tau$, $u_i$ generates $2\cdot|K|+2$ encrypted numbers, where $|K|=c$ is the number of classes.
\par

\begin{algorithm}[t]
\DontPrintSemicolon
\small
\KwIn{${[x]}$: ciphertext, $\mathbb{Z}_q$: secret sharing scheme space \newline
${pk}$: the public key, $\{{sk}_i\}_{i=1}^m$: partial secret keys
}
\KwOut{$\langle x \rangle = ({\langle x \rangle}_1, \cdots, {\langle x \rangle}_m)$: secretly shared $x$}
{
    \For{$i \in [1,m]$} {
        $[r_i] \leftarrow $ $u_i$ randomly chooses $r_i \in \mathbb{Z}_q$ and encrypts it \\
        $u_i$ sends $[r_i]$ to $u_1$
    }
    $u_1$ computes $[e] = [x] \oplus [r_1] \oplus \cdots \oplus [r_m]$ \\ 
    $e \leftarrow$ clients jointly decrypt $[e]$ \\ 
    $u_1$ sets ${\langle x \rangle}_1 = e - r_1 \mod q$ \\
    \For{$i \in [2,m]$}{
        $u_i$ sets ${\langle x \rangle}_i = - r_i \mod q$ 
    }
}
\caption{Conversion to secretly shared value}\label{alg:mpc-conversion}
\end{algorithm}

\vspace{1mm}
\noindent
\textbf{MPC computation.} After the clients generate the encrypted statistics mentioned above (i.e., $[g_{l,k}]$, $[g_{r,k}]$, $[n_l]$, $[n_r]$), they execute an MPC protocol to identify the best split of the current node. 
Towards this end, the clients first invoke 
Algorithm~\ref{alg:mpc-conversion} to convert each encrypted number $[x]$ into a set of secret shares $\{\langle x\rangle_i\}_{i=1}^m$, where $\langle x \rangle_i$ is given to $u_i$. The general idea of Algorithm~\ref{alg:mpc-conversion} is from \cite{CramerDN01,DamgardPSZ12,Zheng2019}. We use $\langle x \rangle$ to denote that the $x$ is secretly shared among the clients.

\begin{algorithm}[t]
\DontPrintSemicolon
\small
\KwIn{$\{D_i\}_{i=1}^m$: local datasets, $\{F_i\}_{i=1}^m$: local features \newline
${Y}$: label set, $[\boldsymbol{\alpha}]$: encrypted mask vector \newline
${pk}$: the public key, $\{ {sk}_i\}_{i=1}^m$: partial secret keys
}
\KwOut{$T$: decision tree model}
{
    \If{prune conditions satisfied} {
        classification: return leaf node with majority class \\
        regression: return leaf node with mean label value
    }
    \Else {
        the super client computes $[\BbbGammaVar]$ and broadcasts it \\
        \For{$i \in [1,m]$} {
            \For{$j \in F_i$}{
                \For{$s \in [1,|S_{ij}|]$}{
                    $u_i$ computes encrypted statistics by $[\BbbGammaVar]$ \\
                }
            }
        }
        clients convert encrypted statistics to shares \\
        determine the best split identifier $(i^*,j^*,s^*)$ \\
        client $i^*$ computes $[\boldsymbol{\alpha}_l], [\boldsymbol{\alpha}_r]$ and broadcasts them \\
        return a tree with $j^*$-th feature and $s^*$-th split value that has two edges, build tree recursively 
    }
}
\caption{Pivot DT training (basic protocol)}\label{alg:private-decision-trees}
\end{algorithm}

After the above conversion, the clients obtain secretly shared statistics $\langle n_l \rangle$, $\langle n_r \rangle$, $\langle g_{l,k} \rangle$, and $\langle g_{r,k} \rangle$ (for each class $k \in K$) for each possible split $\tau$ of the current node. Using these statistics, the clients identify the best split of the current node as follows. 

Consider a split $\tau$ and the two child nodes that it induces. To evaluate the Gini impurity of the left (resp.\ right) child node, the clients need to derive, for each class $k \in K$, the fraction $p_{l, k}$ (resp.\ $p_{r, k}$) of samples on the node that are labeled with $k$. Observe that 
\begin{align}
    p_{l,k} = \frac{ g_{l,k} }{\sum_{k'\in K} g_{l,k'}}, \quad  p_{r,k}  = \frac{ g_{r,k}}{\sum_{k'\in K}  g_{r,k'} }.
    \label{eq:class-probability}
\end{align}
In addition, recall that the clients have obtained, for each class $k \in K$, the secretly shared values $\langle g_{l,k} \rangle$ and $\langle g_{r,k} \rangle$. Therefore, the clients can jointly compute $\langle p_{l, k} \rangle$ and $\langle p_{r, k} \rangle$ using the secure addition and secure division operators in SPDZ (see Section~\ref{subsec:preliminaries:secure-multiparty-computation}), without disclosing $p_{l, k}$ and $p_{r, k}$ to any client. 
With the same approach, the clients use $\langle n_l \rangle$ and $\langle n_r \rangle$ to securely compute $\langle w_l \rangle$ and $\langle w_r \rangle$, where $w_l = \frac{n_l}{ n_l  +  n_r}$ and $w_r = \frac{n_r}{ n_l  +  n_r}$.
Given $\langle p_{l, k} \rangle$, $\langle p_{r, k} \rangle$, $\langle w_l \rangle$, and $\langle w_r \rangle$, the clients can then compute the impurity gain of each split $\tau$ (see Eqn.~\eqref{eq:best-split}) in secretly shared form, using the secure addition and secure multiplication operators in SPDZ. 

Finally, the clients jointly determine the best split using a secure maximum computation as follows. First, each client $u_i$ assigns an identifier $(i, j, s)$ to the $s$-th split on the $j$-th feature that she holds. Next, the clients initialize four secretly shared values $\langle \text{\it gain}_{\text{\it max}} \rangle$, $\langle i^* \rangle$, $\langle j^* \rangle$, $\langle s^* \rangle$, all with $\langle -1 \rangle$.
After that, they will compare the secretly shared impurity gains of all splits, and securely record the identifier and impurity gain of the best split in $\langle i^* \rangle$, $\langle j^* \rangle$, $\langle s^* \rangle$, and $\langle \text{\it gain}_{\text{\it max}} \rangle$, respectively.
Specifically, for each split $\tau$, the clients compare its impurity gain $\langle \text{\it gain}_\tau \rangle$ with $\langle \text{\it gain}_\text{\it max} \rangle$ using secure comparison (see Section~\ref{subsec:preliminaries:secure-multiparty-computation}). Let $\langle \text{\it sign} \rangle$ be the result of the secure comparison, i.e., $\text{\it sign} = 1$ if $\text{\it gain}_\tau > \text{\it gain}_\text{\it max}$, and $\text{\it sign} = 0$ otherwise. Then, the clients securely update $\langle \text{\it gain}_\text{\it max} \rangle$ using the secretly shared values, such that
$\text{\it gain}_\text{\it max} = \text{\it gain}_{\text{\it max}} \cdot (1 - \text{\it sign}) + \text{\it gain}_\tau \cdot \text{\it sign}$. 
The best split identifier is updated in the same manner. After examining all splits, the clients obtain the secretly shared best split identifier $(\langle i^* \rangle, \langle j^* \rangle, \langle s^* \rangle)$.
\color{black}

\vspace{1mm}
\noindent
\textbf{Model update.} Recall that in the basic protocol, the tree model can be released in plaintext. 
Therefore, the clients reconstruct the secretly shared identifier. 
Then, the $i^*$-th client can retrieve the two indicator vectors $\boldsymbol{v}_l$ and $\boldsymbol{v}_r$ for the $s^*$-th split of the $j^*$-th feature. 
After that, she executes element-wise homomorphic multiplication on the two vectors by $[\boldsymbol{\alpha}]$, obtaining $[\boldsymbol{\alpha}_l]$ and $[\boldsymbol{\alpha}_r]$ for the two branches, and broadcasts them to the other clients. 
Note that $[\boldsymbol{\alpha}_l]$ and $[\boldsymbol{\alpha}_r]$ exactly specify the available samples on the two child nodes, respectively. 
For example, in Figure~\ref{fig:classification-example}, if the current split of the `deposit' feature is selected, $u_3$ can compute $[\boldsymbol{\alpha}_l] = ([1],[1],[0],[0],[0])$ using $\boldsymbol{v}_l$ and $[\boldsymbol{\alpha}]$, indicating that Samples 1 and 2 are available on the left child node.

\subsection{Regression Tree Training}\label{subsec:regression-train}

For the regression tree, since the label is continuous, the central part is to compute the label variance. The MPC computation step and model update step are similar to that of classification, thus, we only present the difference in the local computation step.

According to the label variance formula Eqn.~\eqref{eq:variance}, the super client can construct two auxiliary vectors $\boldsymbol{\beta}_1 = (y_1, \cdots, y_n)$ and $\boldsymbol{\beta}_2 = (y_1^2, \cdots, y_n^2)$, where the elements in $\boldsymbol{\beta}_1$ are the original training labels while the elements in $\boldsymbol{\beta}_2$ are the square of the original training labels. Next, she computes element-wise homomorphic multiplication on $\boldsymbol{\beta}_1$ (resp. $\boldsymbol{\beta}_2$) by $[\boldsymbol{\alpha}]$, obtaining $[\boldsymbol{\gamma}_1]$ (resp. $[\boldsymbol{\gamma}_2]$). Then, she broadcasts $[\BbbGammaVar] = \{[\boldsymbol{\gamma}_1], [\boldsymbol{\gamma}_2]\}$ to all clients.
Similarly, each client computes the following encrypted statistics for any local split: 
\begin{align}
    [n_l] = \boldsymbol{v}_l \odot [\boldsymbol{\alpha}], \quad [g_{l,1}] = \boldsymbol{v}_l \odot [\boldsymbol{\gamma}_1], \quad [g_{l,2}] = \boldsymbol{v}_l \odot [\boldsymbol{\gamma}_2]
    \label{eq:regression-statistics}
\end{align}
where $[n_l], [g_{l,1}], [g_{l,2}]$ are the encrypted number of samples, and the encrypted sum of $[\boldsymbol{\gamma}_1]$ and $[\boldsymbol{\gamma}_2]$ of the available samples, for the left branch. Similarly, these encrypted statistics can be converted into secretly shared values and the best split identifier can be decided based on Eqn.~\eqref{eq:variance}.

\begin{figure}[t]
\begin{subfigure}{.62\columnwidth}
  \centering
  \includegraphics[width=0.95\columnwidth]{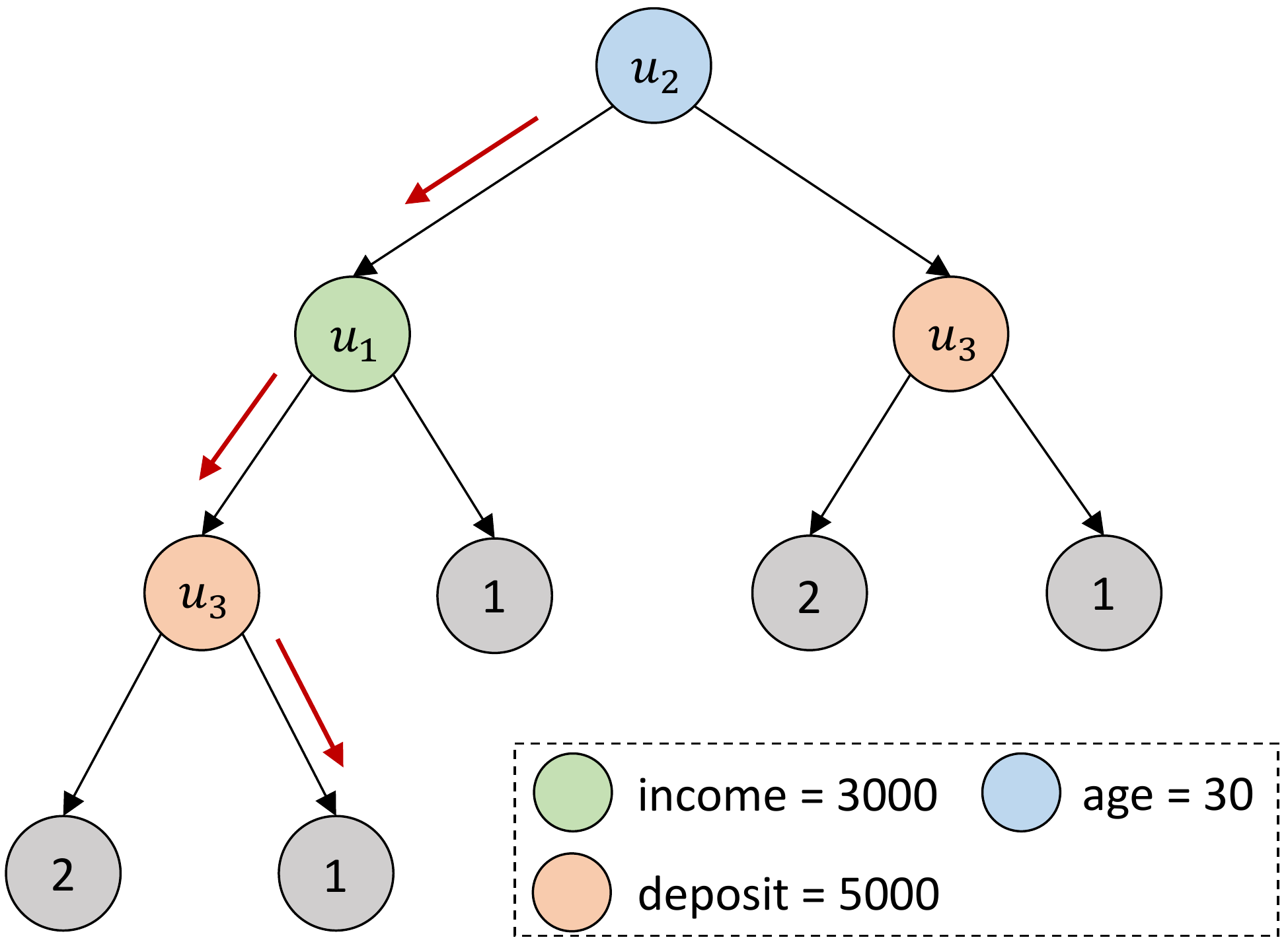}  
  \caption{Tree model}
  \label{fig:sub-tree-first}
\end{subfigure}
\begin{subfigure}{.36\columnwidth}
  \centering
  \includegraphics[width=0.95\columnwidth]{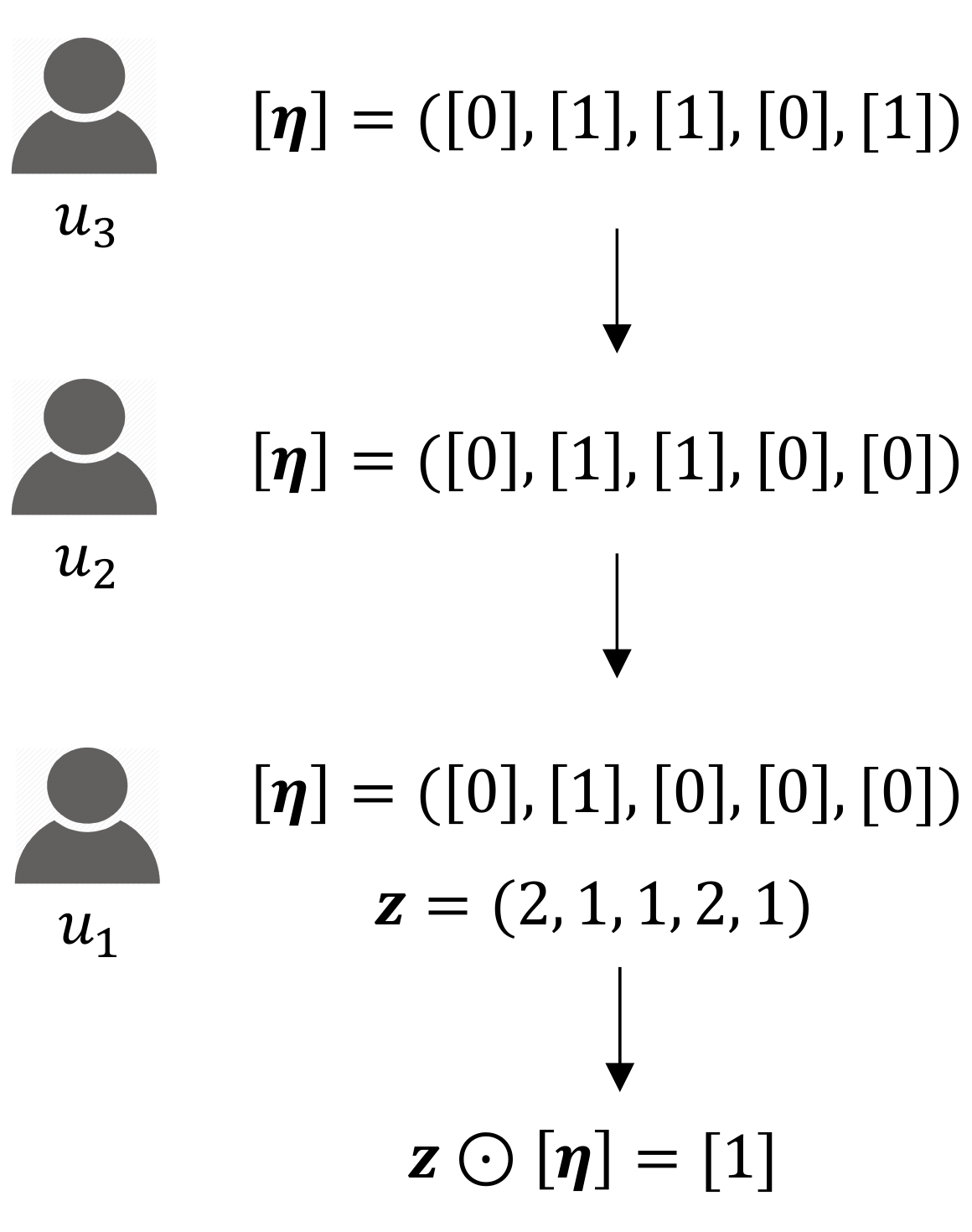}  
  \caption{{Prediction}}
  \label{fig:sub-tree-second}
\end{subfigure}
\vspace{-2mm}
\caption{Tree model prediction example with a sample $(\text{age}=25, \text{income}=2500, \text{deposit}=6000)$: (a) tree model: colored circles denote internal nodes and gray circles denote leaf nodes; (b) prediction: clients update a encrypted prediction vector in a round-robin manner. }
\label{fig:tree-prediction}
\end{figure}

Algorithm~\ref{alg:private-decision-trees} describes the privacy preserving decision tree training protocol. {Lines 1-3 check the pruning conditions and compute the leaf label if any condition is satisfied.} 
Note that with the encrypted statistics, these executions can be easily achieved by secure computations. 
Lines 5-13 find the best split and build the tree recursively, where
lines 5-9 are the local computation step for computing encrypted split statistics; line 10-11 are the MPC computation step that converts the encrypted statistics into secretly shared values and decides the best split identifier using MPC; and line 12 is the model update step, which computes the encrypted indicator vectors for the child nodes given the best split. 

\subsection{Tree Model Prediction}\label{subsec:prediction-trees}

After releasing the plaintext tree model, the clients can jointly make a prediction given a sample. In vertical FL, the features of a sample are distributed among the clients. 
Figure~\ref{fig:sub-tree-first} shows an example of a released model, where each internal node represents a feature with a split threshold owned by a client, and each leaf node represents a predicted label on that path. To predict a sample, a naive method is to let the super client coordinate the prediction process \cite{ChengCorr19}: starting from the root node, the client who has the node feature compares its value with the split threshold, and notifies super client the next branch; then the prediction is forwarded to the next node until a leaf node is reached. However, this method discloses the prediction path, from which a client can infer the other client's feature value along that path.

To ensure that no additional information {other than the predicted output} is leaked, we propose a distributed prediction method, as shown in Algorithm~\ref{alg:private-decision-trees-prediction}. Let $\boldsymbol{z} = (z_1, \cdots, z_{t+1})$ be the \textit{leaf label vector} of the leaf nodes in the tree model, where $t$ is the number of internal nodes. Note that all clients know $\boldsymbol{z}$ since the tree model is public in this protocol. Given a sample, clients collaborate to update an
encrypted prediction vector $[\boldsymbol{\eta}] = ([1],\cdots, [1])$ with size $t+1$ in a round-robin manner. Each element in $[\boldsymbol{\eta}]$ indicates if a prediction path is possible with encrypted form. \par

Without loss of generality, we assume that the prediction starts with $u_m$ and ends with $u_1$. If a prediction path is possible from the perspective of a client, then the client multiplies the designated element in $[\boldsymbol{\eta}]$ by 1 using homomorphic multiplication, otherwise by 0. 
Figure~\ref{fig:sub-tree-second} illustrates an example of this method. Starting from $u_3$, given the feature value `deposit = 6000', $u_3$ initializes $[\boldsymbol{\eta}]$ and updates it to $([0],[1],[1],[0],[1])$, since she can eliminate the first and fourth prediction paths after comparing her value with the split threshold `deposit = 5000'. Then, $u_3$ sends $[\boldsymbol{\eta}]$ to the next client for updates. 
After all clients' updates, there is only one $[1]$ in $[\boldsymbol{\eta}]$, which indicates the true prediction path.
Finally, $u_1$ computes $\boldsymbol{z} \odot [\boldsymbol{\eta}]$ to get the encrypted prediction output, and decrypts it jointly with all clients.

\subsection{Security Guarantees}\label{subsec:privacy-analysis}
\newtheorem{theorem}{Theorem}
\begin{theorem}
The basic protocol of Pivot securely realizes the ideal functionalities $\mathcal{F}_{\text{DTT}}$ and $\mathcal{F}_{\text{DTP}}$ against a semi-honest adversary who can statically corrupt up to $m-1$ out of $m$ clients.
\label{theorem:ppdt}
\vspace{-0.2cm}
\end{theorem}

\noindent
\textit{Proof Sketch.}
We need to show that, in the view of an adversary $\mathcal{A}$, any information learned by the protocol can be learned directly from the input it has corrupted and the output it receives. \par 

For model training, the proof can be reduced to the computations on one tree node because each node can be computed separately given that its output is public \cite{LindellP00, LindellP09}. There are two cases. First, when a given node is an internal node: (i) if the super client is corrupted, nothing is revealed in the local computation step regarding the honest client's data; while the MPC conversion \cite{CramerDN01} and additive secret sharing scheme \cite{DamgardPSZ12} are secure, thus, the MPC computation step is secure; finally, in the model update step, if $i^*$ is an honest client, the transmitted message $[\boldsymbol{\alpha}]$ is secure for the threshold Paillier scheme \cite{Paillier99} is secure. (ii) if the super client is not corrupted, the only difference is the transmitted encrypted label information $[\BbbGammaVar]$, which is also secure. Second, when a given node is a leaf node: (i) if the super client is corrupted, nothing is revealed since the honest client does not have the labels; (ii) if the super client is not corrupted, the transmitted messages are the encrypted sample number of each class (for classification) and the encrypted mean label (for regression), which are secure. Therefore, $\mathcal{A}$ learns no additional information from the protocol execution, the security follows. 

For model prediction, the adversary $\mathcal{A}$ views an encrypted prediction vector $[\boldsymbol{\eta}]$ updated by the honest client(s) and the encrypted prediction output $[\bar{k}]$, thus, no more information is learned (the decrypted prediction output is public) for the threshold Paillier scheme is secure. 
$\Box$

\begin{algorithm}[t!]
\DontPrintSemicolon
\small
\KwIn{$T$: decision tree model, $\{\textbf{x}_i\}_{i=1}^m$: input sample \newline
${pk}$: the public key, $\{ {sk_i}\}_{i=1}^m$: partial secret keys
}
\KwOut{$\bar{k}$: predicted label}
{
    \For{$i \in [m,1]$} {
        \If{$i == m$}{
            $u_i$ initializes $[\boldsymbol{\eta}] = ([1], \cdots, [1])$ with size $t+1$ \\
        }
        \If{$i > 1$}{
            $u_i$ updates $[\boldsymbol{\eta}]$ using ($T$, $\boldsymbol{x}_i$) \\
            $u_i$ sends $[\boldsymbol{\eta}]$ to $u_{i-1}$ \\
        }
        \Else{
            $u_i$ updates $[\boldsymbol{\eta}]$ using ($T$, $\boldsymbol{x}_i$) \\
            $u_i$ initializes label vector $\boldsymbol{z} = (z_1, \cdots, z_{t+1})$ \\
            $u_i$ computes $[\bar{k}] = \boldsymbol{z} \odot [\boldsymbol{\eta}]$ \\
        }
    }
    clients jointly decrypt $[\bar{k}]$ by $\{sk_i\}_{i=1}^m$ and return $\bar{k}$ 
}
\caption{Pivot DT prediction (basic protocol)}\label{alg:private-decision-trees-prediction}
\end{algorithm}

%% file: tex-files/sec-enhanced-solution.tex
\section{Enhanced Protocol}\label{sec:enhanced-solution}

The basic protocol guarantees that no intermediate information is disclosed. However, after obtaining the public model, colluding clients may extract private information of a target client's training dataset, with the help of their own datasets.
We first present two possible privacy leakages in Section~\ref{subsec:privacy-leakages} and then propose an enhanced protocol that mitigates this problem by concealing some model information in Section~\ref{subsec:hiding-split-label}. 
The security analysis is given in Section~\ref{subsec:enhanced-privacy-analysis}.

\subsection{Privacy Leakages}\label{subsec:privacy-leakages}

We identify two possible privacy leakages: the training label leakage and the feature value leakage, regarding a target client's training dataset. The intuition behind the leakages is that the colluding clients are able to split the sample set based on the split information in the model and their own datasets. We illustrate them by the following two examples given the tree model in Figure~\ref{fig:tree-prediction}. 

\vspace{-0.15cm}
\newdef{example}{Example}
\begin{example}
(Training label leakage). Assume that $u_2$ and $u_3$ collude, let us see the right branch of the root node. $u_2$ knows exactly the sample set in this branch, say $D_{\text{age $>$ 30}}$, as all samples are available on the root node and he can just split his local samples based on `age = 30'. Then, $u_3$ can classify this set into two subsets given the `deposit=5000' split, say $D_{\text{age $>$ 30} \bigwedge \text{deposit $\leq$ 5000}}$ and $D_{\text{age $>$ 30} \bigwedge \text{deposit $>$ 5000}}$, respectively. Consequently, according to the plaintext class labels on the two leaf nodes, colluding clients may infer that the samples in $D_{\text{age $>$ 30} \bigwedge \text{deposit $\leq$ 5000}}$ are with class 2 and vise versa, with high probability. 
\label{example:label-leakage-training}
\vspace{-0.15cm}
\end{example}
\begin{example}
(Feature value leakage). Assume that $u_1$ and $u_2$ collude, let us see the path of $u_2 \rightarrow u_1 \rightarrow u_3$ (with red arrows). Similar to Example~\ref{example:label-leakage-training}, $u_1$ and $u_2$ can exactly know the training sample set on the `$u_3$' node before splitting, say $D'$. In addition, recall that $u_1$ is the super client who has all sample labels, thus, he can easily classify ${D}'$ into two sets by class, say ${D}_1'$ and ${D}_2'$, respectively. Consequently, the colluding clients may infer that the samples in ${D}_2'$ have `deposit $\leq$ 5000' and vise versa, with high probability.
\label{example:feature-leakage-training}
\vspace{-0.15cm}
\end{example}

Note that these two leakages happen when the clients (except the target client) along a path collude. 
Essentially, given the model, the colluding clients (without super client) may infer labels of some samples in the training dataset if there is no feature belongs to the super client along a tree path; similarly, if the super client involves in collusion, the feature values of some samples in the training dataset of a target client may be inferred. 

\subsection{Hiding Label and Split Threshold}\label{subsec:hiding-split-label}

Our observation is that these privacy leakages can be mitigated if the split thresholds on the internal nodes and the leaf labels on the leaf nodes in the model are concealed from all clients.
Without such information, the colluding clients can neither determine how to split the sample set nor what leaf label a path owns. We now discuss how to hide these information in the model.

For the leaf label on each leaf node, the clients can convert it to an encrypted value, instead of reconstructing its plaintext. Specifically, after obtaining the secretly shared leaf label (e.g., $\langle k \rangle$) using secure computations (Lines 1-3 in Algorithm~\ref{alg:private-decision-trees}), each client encrypts her own share of $\langle k \rangle$ and broadcasts to all clients. Then, the encrypted leaf label can be computed by summing up these encrypted shares using homomorphic addition. As such, the leaf label is concealed.

For the split threshold on each internal node, the clients hide it by two additional computations in the model update step.
Recall that in the basic protocol, the best split identifier $(\langle i^* \rangle, \langle j^* \rangle, \langle s^* \rangle)$ is revealed to all clients after the MPC computation in each iteration. 
In the enhanced protocol, we assume that $\langle s^* \rangle$ is not revealed, thus the split threshold can be concealed. 
To support the tree model update without disclosing $s^*$ to the $i^*$-th client, we first use the private information retrieval (PIR) \cite{WuWZLCL18, Wu2019} technique to privately select the split indicator vectors of $s^*$.

\vspace{1mm}
\noindent
\textbf{Private split selection.} Let $n' = |S_{ij}|$ denote the number of splits of the $j^*$-th feature of the $i^*$-th client. We assume $n'$ is public for simplicity. Note that the clients  can further protect $n'$ by padding placeholders to a pre-defined threshold number. Instead of revealing $\langle s^* \rangle$ to the $i^*$-th client, 
the clients jointly convert $\langle s^* \rangle$ into an encrypted indicator vector $[\boldsymbol{\lambda}] = ([\lambda_1], \cdots, [\lambda_{n'}])^T$, such that $\lambda_t = 1$ when $t = s^*$ and $\lambda_t = 0$ otherwise, where $t \in \{1, \cdots, n'\}$. This vector is sent to the $i^*$-th client for private split selection at local. Let ${\boldsymbol{V}}^{n \times n'} = (\boldsymbol{v}_{1}, \cdots, \boldsymbol{v}_{n'})$ be the split indicator matrix, where $\boldsymbol{v}_{t}$ is the split indicator vector of the $t$-th split of the $j^*$-th feature (see Section~\ref{subsec:classification-train}). The following theorem \cite{Wu2019} suggests that the $i^*$-th client can compute the encrypted indicator vector for the $s^*$-th split without disclosing $s^*$.
\begin{theorem}
Given an encrypted indicator vector $[\boldsymbol{\lambda}] = ([\lambda_1], \cdots, [\lambda_{n'}])^T$ such that $[\lambda_{s^*}] = [1]$ and $[\lambda_{t}] = [0]$ for all $t \neq s^*$, and the indicator matrix ${\boldsymbol{V}}^{n \times n'} = (\boldsymbol{v}_{1}, \cdots, \boldsymbol{v}_{n'})$, then $[\boldsymbol{v}_{s^*}] = \boldsymbol{V} \bigotimes [\boldsymbol{\lambda}]$. $\Box$
\label{theorem:ref-private-selection}
\end{theorem}
The notion $\bigotimes$ represents the homomorphic matrix multiplication, which executes homomorphic dot product operations between each row in $\boldsymbol{V}$ and $[\boldsymbol{\lambda}]$. We refer the interested readers to \cite{Wu2019} for the details. 
\par
 
For simplicity, we denote the selected $[\boldsymbol{v}_{s^*}]$ as $[\boldsymbol{v}]$. The encrypted split threshold can also be obtained by homomorphic dot product between the encrypted indicator vector $[\boldsymbol{\lambda}]$ and the plaintext split value vector of the $j^*$-th feature.

\vspace{1mm}
\noindent
\textbf{Encrypted mask vector updating.} After finding the encrypted split vector $[\boldsymbol{v}]$, we need to update the encrypted mask vector $[\boldsymbol{\alpha}]$ for protecting the sample set recursively. This requires element-wise multiplication between $[\boldsymbol{\alpha}]$ and $[\boldsymbol{v}]$.  
Thanks to the MPC conversion algorithm, we can compute $[\boldsymbol{\alpha}] \cdot [\boldsymbol{v}]$ as follows \cite{CramerDN01}. For each element pair $[\alpha_j]$ and $[v_{j}]$ where $j \in [1,n]$, we first convert $[\alpha_j]$ into $\langle \alpha_j \rangle = ({\langle \alpha_j \rangle}_1, \cdots, {\langle \alpha_j \rangle}_m)$ using Algorithm~\ref{alg:mpc-conversion}, where ${\langle \alpha_j \rangle}_i (i\in \{1,\cdots,m\})$ is the share hold by $u_i$; then each client $u_i$ executes homomorphic multiplication ${\langle \alpha_j \rangle}_i \otimes [v_j] = [{\langle \alpha_j \rangle}_i \cdot v_j]$ and sends the result to the $i^*$-th client; finally, the $i^*$-th client can sum up the results using homomorphic addition:
\begin{align}
    [\alpha_j'] & = [{\langle \alpha_j \rangle}_1 \cdot v_j] \oplus \cdots \oplus [{\langle \alpha_j \rangle}_m \cdot v_j] = [\alpha_j \cdot v_j] 
    \label{eq:multiplication-encrypted-vectors}
\end{align}
After updating $[\boldsymbol{\alpha}]$, the tree can also be built recursively, similar to the basic protocol. 

\begin{table*}[t]
\centering
\small
\caption{Theoretical analysis}
\vspace{-0.2cm}
\begin{tabular}{ l | l | l }
\toprule
&  Pivot basic protocol & Pivot enhanced protocol \\
\toprule
{Model training} & {$O(n c \bar{d} {b} t) C_{e} + O(c {d} b t) (C_{d} + C_{s}) + O(d b t) C_{c}$} & {$O(n c \bar{d} {b} t) C_{e} + O(c{d}b t + n t) C_{d}+ O(c d {b} t) C_{s} + O(db t) C_{c}$} \\ 
-- local computation & $O(n c \bar{d} {b}) C_{e}$ & $O(n c \bar{d} {b}) C_{e}$ \\
-- mpc computation & $O(c {d} b) (C_{d} + C_{s}) + O(db)C_c$ & $O(c {d} b) (C_{d} + C_{s}) + O(db)C_c$ \\
-- model update & $O(n) C_{e}$ & $O(n b) C_{e} + O(n)C_d$ \\
\midrule
Model prediction & $O(m t) C_{e} + O(1) C_d$ & $O(t) (C_{s} + C_c)$ \\
\bottomrule
\end{tabular}
\label{table:performace-analysis}
\vspace{2mm}
\end{table*}

\vspace{1mm}
\noindent    
\textbf{Secret sharing based model prediction.} The prediction method in the basic protocol is not applicable here as the clients cannot directly compare their feature values with the encrypted split thresholds. Hence, the clients first convert the encrypted split thresholds and encrypted leaf labels into secretly shared form and make predictions on the secretly shared model using MPC. Let $\langle \boldsymbol{z} \rangle$ with size $(t+1)$ denote the secretly shared leaf label vector, where $t$ is the number of internal nodes.

To make the prediction given a sample, the clients also provide the distributed feature values in secretly shared form. 
Similar to the prediction in the basic protocol, the clients initialize a secretly shared prediction vector $\langle \boldsymbol{\eta} \rangle$ with size $(t+1)$, indicating if a prediction path is possible. Then, they compute this vector as follows.

The clients initialize a secretly shared marker $\langle 1 \rangle$ for the root node. 
Starting from root node, the clients recursively compute the markers of its child nodes until all leaf nodes are reached. Then, the marker of each leaf node is assigned to the corresponding position in $\langle \boldsymbol{\eta} \rangle$, and there is only one $\langle 1 \rangle$ element in $\langle \boldsymbol{\eta} \rangle$, specifying the real prediction path in a secret manner.
Specifically, each marker is computed by secure multiplication between its parent node's marker and a secure comparison result (between the secretly shared feature value and split threshold on this node).
For example, in Figure~\ref{fig:sub-tree-first}, the split threshold on the root node will be $\langle 30 \rangle$ while the feature value will be $\langle 25 \rangle$, then $\langle 1 \rangle$ is assigned to its left child and $\langle 0 \rangle$ to its right child. The clients know nothing about the assigned markers due to the computations are secure. 
Finally, the secretly shared prediction output can be computed easily by a dot product between $\langle \boldsymbol{z} \rangle$ and $\langle \boldsymbol{\eta} \rangle$, using secure computations.
    
\vspace{1mm}
\noindent
\textbf{Discussion.} A noteworthy aspect is that the clients can also choose to hide the feature $\langle j^* \rangle$ by defining $n'$ as the total number of splits on the $i^*$-th client, or even the client $\langle i^* \rangle$ that has the best feature by defining $n'$ as the total number of splits among all clients. By doing so, the leakages could be further alleviated. However, the efficiency and interpretability would be degraded greatly. In fact, there is a trade-off between privacy and efficiency (interpretability) for the released model. The less information the model reveals, the higher privacy while the lower efficiency and less interpretability the clients obtain, and vise versa. \par

\subsection{Security Guarantees}\label{subsec:enhanced-privacy-analysis}
\begin{theorem}
The enhanced protocol of Pivot securely realizes the ideal functionalities $\mathcal{F}_{\text{DTT}}$ and $\mathcal{F}_{\text{DTP}}$ against a semi-honest adversary who can statically corrupt up to $m-1$ out of $m$ clients.
\label{theorem:ppdt-enhanced}
\end{theorem}
\vspace{-0.2cm}

\noindent
\textit{Proof Sketch.} For model training, 
the only difference from the basic protocol is the two additional computations (private split selection and encrypted mask vector updating) in the model update step, which are computed using threshold Paillier scheme and MPC conversion. Thus, the security follows. For model prediction, since the additive secret sharing scheme is secure and the clients compute a secretly shared marker for every possible path, the adversary learns nothing except the final prediction output.
$\Box$

%% file: tex-files/sec-analysis.tex
\section{Theoretical Analysis}\label{sec:performance-analysis}

We theoretically analyze the Pivot basic protocol and Pivot enhanced protocol in terms of computational cost for model training and model prediction, as summarized in Table~\ref{table:performace-analysis}. 
Let $C_{e}$ and $C_{s}$ roughly denote the costs for computations on a homomorphic encrypted value and on a secretly shared value, respectively. Due to that the threshold decryption (involving decryption of each client and combination via network communication) and secure comparison (involving multi-round network communications among the clients) are more time-consuming than the other computations, we consider these two operations separately for better analysis, and denote the costs of them by $C_{d}$ and $C_{c}$, respectively. Let $\bar{d} = \max(\{d_i\}_{i=1}^m)$ be the maximum number of features any client holds, $b$ be the maximum number of splits any feature has, and $c$ be the number of classes.

\vspace{1mm}
\noindent
\textbf{Model training.} With the basic protocol, the computational cost of a client in each iteration includes: (i) local computation step: the encrypted label vectors computed by the super client, i.e., $O(nc) C_{e}$, and the encrypted statistics computed by the clients, i.e., $O(n c \bar{d} b) C_{e}$, where $\bar{d} b$ is number of local splits; (ii) MPC computation step: the MPC conversion for encrypted statistics of total splits, i.e., $O(c d b) C_{d}$, and the best split determined using $O(c d b)$ statistics, i.e., $O(c d b) C_{s} + O(d b) C_{c}$, where $db$ is the number of total splits; and (iii) model update step: the update of encrypted mask vectors, i.e., $O(n) C_{e}$.
Thus, the total cost is
$O(n c \bar{d} {b} t) C_{e} + O(c {d} b t) (C_{d} + C_{s}) + O(d b t) C_{c}$, where $t$ is the number of internal nodes in the tree model. 
With the enhanced protocol, the only difference is the two additional computations in the model update step: private split selection on ${b}$ split indicator vectors, i.e., $O(n {b}) C_{e}$, and encrypted mask vector update that mainly requires $n$ threshold decryption operations, i.e., $O(n) C_{d}$. 
{Thus, the total cost is $O(n c \bar{d} {b} t) C_{e} + O(c{d}b t + nt) C_{d} + O(c d {b} t) C_{s} + O(db t) C_{c}$.} \par

\vspace{1mm}
\noindent
\textbf{Model prediction.} With the basic protocol, the computational cost of prediction is updating an encrypted prediction vector with size $(t+1)$ in a Robin round, i.e., $O(m t) C_{e}$, the homomorphic dot product between the encrypted prediction vector and the plaintext label vector, i.e., $O(t) C_{e}$, and the threshold decryption of the final prediction output, i.e., $O(1) C_{d}$. Thus, the total cost is $O(m t) C_{e} + O(1) C_d$. With the enhanced protocol, the computational cost includes the secure comparison of $t$ internal nodes and the secure dot product between the prediction vector and the label vector, i.e., $O(t) (C_{s} + C_c)$.

In summary, regarding model training, the computational cost of the enhanced protocol is always larger than that of the basic protocol because the two additional computations are extra costs. Regarding model prediction, whether the basic protocol is better depends on the number of clients $m$ and the relationship between ciphertext computation cost and secure computation cost. We will experimentally evaluate the two protocols in Section \ref{sec:experiments}. 

%% file: tex-files/sec-extensions.tex
\section{Extensions to OTHER ML MODELS}\label{sec:extensions}

So far, Pivot supports a single tree model. Now we briefly present how to extend the basic protocol to ensemble tree models, including random forest (RF) \cite{Breiman01} and gradient boosting decision tree (GBDT) \cite{Friedman00, FuJSC19} {in Section \ref{subsec:random-forest} and Section \ref{subsec:gbdt}, respectively.} Same as the basic protocol, we assume that all the trees can be released in plaintext. {The extension to other machine learning models is discussed in Section \ref{subsec:extension-other-models}.}

\subsection{Random Forest}\label{subsec:random-forest}

RF constructs a set of independent decision trees in the training stage and outputs the class that is the mode of the classes (for classification) or mean prediction (for regression) of those trees in the prediction stage. \par

For model training, the extension from a single decision tree is natural since each tree can be built (using Algorithm \ref{alg:private-decision-trees}) and released separately. 
For model prediction, after obtaining the encrypted predicted label of each tree, the clients can easily convert these encrypted labels into secret shares for majority voting using secure maximum computation (for classification) or compute the encrypted mean prediction by homomorphic computations (for regression).

\subsection{Gradient Boosting Decision Trees}\label{subsec:gbdt}

GBDT uses decision trees as weak learners and improves model quality with a boosting strategy \cite{Friedman98}. The trees are built sequentially where the training labels for the next tree are the prediction losses between the ground truth labels and the prediction outputs of previous trees. 

\vspace{1mm}
\noindent
\textbf{Model training.} The extension to GBDT is non-trivial, since we need to prevent the super client from knowing the training labels of each tree except the first tree (i.e., intermediate information) while facilitating the training process. 

We first consider GBDT regression. Let $W$ be the number of rounds and a regression tree is built in each round.
Let ${Y}^{w}$ be the training label vector of the $w$-th tree. 
We aim to protect ${Y}^{w}$ by keeping it in an encrypted form. 
After building the $w$-th tree where $w \in \{1, \cdots, W-1\}$, the clients jointly make predictions for all training samples to get an encrypted estimation vector $[\bar{{Y}}^{w}]$; 
then the clients can compute the encrypted training labels $[{Y}^{w+1}]$ of the $(w+1)$-th tree given $[{Y}^{w}]$ and $[\bar{{Y}}^{w}]$. 
Besides, note that in Section~\ref{subsec:regression-train}, an encrypted label square vector $[\boldsymbol{\gamma}_2^{w+1}]$ is needed, which is computed by element-wise homomorphic multiplication between $\boldsymbol{\beta}_2^{w+1}$ and $[\boldsymbol{\alpha}]$. However, $\boldsymbol{\beta}_2^{w+1}$ is not plaintext here since the training labels are ciphertexts. Thus, the clients need expensive element-wise ciphertext multiplications (see Section~\ref{subsec:hiding-split-label}) between $[\boldsymbol{\beta}_2^{w+1}]$ and $[\boldsymbol{\alpha}]$ in each iteration. 
To optimize this computation, we slightly modify our basic protocol.
Instead of letting the super client compute $[\boldsymbol{\gamma}_2^{w+1}]$ in each iteration, we now let the client who has the best split update $[\boldsymbol{\gamma}_2^{w+1}]$ along with $[\boldsymbol{\alpha}]$ using the same split indicator vector and broadcast them to all clients. In this way, the clients only need to compute $[{\boldsymbol{\gamma}}_2^{w+1}]$ using $[\boldsymbol{\beta}_2^{w+1}]$ and $[\boldsymbol{\alpha}]$ once at the beginning of each round, which reduces the cost.

For GBDT classification, we use the one-vs-the-rest technique by combining a set of binary classifiers. Essentially, the clients need to build a GBDT regression forest for each class, resulting in $W * c$ regression trees in total ($c$ is the number of classes). After each round in the training stage, the clients obtain $c$ trees; and for each training sample, the clients make a prediction on each tree, resulting in $c$ encrypted prediction outputs. Then, the clients jointly convert them into secretly shared values for computing \textit{secure softmax} (which can be constructed using secure exponential, secure addition, and secure division, as mentioned in Section~\ref{subsec:preliminaries:secure-multiparty-computation}), and convert them back into an encrypted form as encrypted estimations. The rest of the computation is the same as regression.

\vspace{1mm}
\noindent
\textbf{Model prediction.} For GBDT regression, the prediction output can be decrypted after homomorphic aggregating the encrypted predictions of all trees. For GBDT classification, the encrypted prediction for each class is the same as that for regression; then the clients jointly convert these encrypted results into secretly shared values for deciding the final prediction output by \textit{secure softmax} function.

\subsection{Other Machine Learning Models}\label{subsec:extension-other-models}

Though we consider tree-based models in this paper, the proposed solution can be easily adopted in other vertical FL models, such as logistic regression (LR), neural networks, and so on. 
The rationale is that these models can often be partitioned into the three steps described in Section~\ref{subsec:classification-train}. 
As a result, the TPHE primitives, conversion algorithm, and secure computation operators can be re-used. 
\par

For example, the clients can train a vertical LR model as follows. 
To protect the intermediate weights of the LR model during the training, the clients initialize an encrypted weight vector, $[\boldsymbol{\theta}] = ([\boldsymbol{\theta}_1], \cdots, [\boldsymbol{\theta}_m])$, where $[\boldsymbol{\theta}_i]$ corresponds to the encrypted weights of features held by client $i$.
In each iteration, for a Sample $t$, each client $i$ first locally aggregates an encrypted partial sum, say $[\xi_{it}]$, by homomorphic dot product between $[\boldsymbol{\theta}_i]$ and Sample $t$'s local features $\textbf{x}_{it}$. Then the clients jointly convert $\{[\xi_{it}]\}_{i=1}^m$ into secretly shared values using Algorithm \ref{alg:mpc-conversion}, and securely aggregate them before computing the secure logistic function. Meanwhile, the super client also provides Sample $t$'s label as a secretly shared value, such that the clients can jointly compute the secretly shared loss of Sample $t$. After that, the clients convert the loss back into the encrypted form (see Section \ref{subsec:hiding-split-label}), and each client can update her encrypted weights $[\boldsymbol{\theta}_i]$ using homomorphic properties, without knowing the loss. 
Besides, the model prediction is a half component of one iteration in training, which can be easily computed.

%% file: tex-files/sec-experiments.tex
\section{Experiments}\label{sec:experiments}

We evaluate the performance of Pivot basic protocol (Section \ref{sec:decision-trees}) and Pivot enhanced protocol (Section \ref{sec:enhanced-solution}) on the decision tree model, as well as the ensemble extensions (Section \ref{sec:extensions}). We present the accuracy evaluation in Section \ref{subsec:experiments-accuracy} and the efficiency evaluation in Section \ref{subsec:experiments-efficiency}. \par

We implement Pivot in C++ and employ the \textit{GMP}\footnote{\url{http://gmplib.org}} library for big integer computation and the \textit{libhcs}\footnote{\url{https://github.com/tiehuis/libhcs}} library for operations of the threshold Paillier scheme.
We utilize the \textit{SPDZ}\footnote{\url{https://github.com/data61/MP-SPDZ}} library for semi-honest additive secret sharing computations. Besides, we apply the \textit{libscapi}\footnote{\url{https://github.com/cryptobiu/libscapi}} library to provide network communications among clients. 
Since the cryptographic primitives only support big integer computations, we convert the floating point datasets into fixed-point integer representation.

\subsection{Experimental Setup}\label{subsec:experiments-setup}

We conduct experiments on a cluster of machines in a local area network (LAN). Each machine is equipped with Intel (R) Xeon (R) CPU E5-1650 v3 @ 3.50GHz$\times$12 and 32GB of RAM, running Ubuntu 16.04 LTS. Unless noted otherwise, the \textit{keysize} of threshold Paillier scheme is 1024 bits and the security parameter of {SPDZ} configuration is 128 bits.

\vspace{1mm}
\noindent
\textbf{Datasets.} We evaluate the model accuracy using three real-world datasets: credit card data (30000 samples with 25 features) \cite{YehL09a}, bank marketing data (4521 samples with 17 features) \cite{MoroCR14}, and appliances energy prediction data (19735 samples with 29 features) \cite{CANDANEDO201781}. The former two datasets are for classification while the third dataset is for regression. \par

We evaluate the efficiency using synthetic datasets, which are generated with \textit{sklearn}\footnote{\url{https://scikit-learn.org/stable/}} library. Specifically, we vary the number of samples ($n$) and the number of total features ($d$) to generate datasets and then equally split these datasets \textit{w.r.t.} features into $m$ partitions, which are held by $m$ clients, respectively. We denote $\bar{d}= d / m$ as the number of features each client holds. For classification tasks, the number of classes is set to 4, and only one client holds the labels.

\begin{table*}[t]
\centering
\small
\caption{Model accuracy comparison with non-private baselines}
\vspace{-0.2cm}
\begin{tabular}{l | l | l | l | l | l | l }
\toprule
Dataset & Pivot-DT & NP-DT & Pivot-RF & NP-RF & Pivot-GBDT & NP-GBDT \\
\toprule
Bank market & 0.886077 & 0.886188 & 0.888619 & 0.890497 & 0.891271 & 0.892044 \\
Credit card &  0.821526 & 0.821533 & 0.823056 & 0.823667 & 0.825167 & 0.827167 \\
Appliances energy & 212.05281 & 211.45229 & 211.55175 & 211.32113 & 211.35326 & 210.75291 \\
\bottomrule
\end{tabular}
\label{table:accuracy-evaluated}
\vspace{2mm}
\end{table*}

\vspace{1mm}
\noindent
\textbf{Baselines.} For accuracy evaluation, we adopt the non-private decision tree (NP-DT), non-private random forest (NP-RF), and non-private gradient-boosting decision tree (NP-GBDT) algorithms from \textit{sklearn} for comparison.
For a fair comparison, we 
adopt the same hyper-parameters for both our protocols and the baselines, e.g., the maximum tree depth, the pruning conditions, the number of trees, etc. \par

For efficiency evaluation, to our knowledge, there is no existing work providing the same privacy guarantee as Pivot.
Therefore, we implement a secret sharing based decision tree algorithm using the SPDZ library (namely, SPDZ-DT) as a baseline. The security parameter of SPDZ-DT is also 128 bits. {Besides, we also implement a non-private distributed decision tree (NPD-DT) algorithm as another baseline to illustrate the overhead of protecting the data privacy. In NPD-DT, the super client broadcasts plaintext labels to all clients, each client computes split statistics and exchanges them in plaintext with others to decide the best split.}

\vspace{1mm}
\noindent
\textbf{Metrics.} For model accuracy, we measure the number of samples that are correctly classified over the total testing samples for classification; and the mean square error (MSE) between the predicted labels and the ground truth labels for regression. 
For efficiency, we measure the total running time of the model training stage and the prediction running time per sample of the model prediction stage.
In all experiments, we report the running time of the online phase because SPDZ did not support the offline time benchmark for the semi-honest additive secret sharing protocol. 

\subsection{Evaluation of Accuracy}\label{subsec:experiments-accuracy}

In terms of accuracy, we compare the performance of the proposed decision tree (Pivot-DT), random forest (Pivot-RF) and gradient boosting decision tree (Pivot-GBDT) algorithms with their non-private baselines on three real world datasets. 
In these experiments, the \textit{keysize} of threshold Paillier scheme is set to 512 bits. We conduct 10 independent trials of each experiment and report the average result.

Table \ref{table:accuracy-evaluated} summarizes the comparison results.
We can notice that the Pivot algorithms achieve accuracy comparable to the non-private baselines. 
There are two reasons for the slight loss of accuracy.
First, we use the fixed-point integer to represent float values, whose precision is thus truncated. 
Second, Pivot has only implemented the basic algorithms, which are not optimized as the adopted baselines.

\begin{table}[b]
\centering
\small
\caption{Parameters adopted in the evaluation}
\vspace{-0.2cm}
\begin{tabular}{l | l | l | l }
\toprule
Parameter & Description & Range & Default \\
\toprule
$m$ & number of clients & {$[2, 10]$} & 3 \\
$n$ & number of samples & $[5K, 200K]$ & $50K$ \\
$\bar{d}$ & number of features & {$[5, 120]$} & $15$ \\
${b}$ & maximum splits & $[2, 32]$ & $8$ \\
$h$ & maximum tree depth & $[2, 6]$ & $4$ \\
$W$ & number of trees & $[2, 32]$ & $-$\\
\bottomrule
\end{tabular}
\label{table:parameters}
\end{table}

\subsection{Evaluation for Efficiency}\label{subsec:experiments-efficiency}

In terms of efficiency, we evaluate the training and prediction time of Pivot with the two protocols (namely Pivot-Basic and Pivot-Enhanced) in Section \ref{subsubsec:evaluation-training} and Section \ref{subsubsec:evaluation-prediction}, by varying the number of clients $(m)$, the number of samples ($n$), the number of features of each client ($\bar{d}$), the maximum number of splits (${b}$), the maximum tree depth ($h$), and the number of trees ($W$) for ensemble methods.

We employ parallelism for threshold decryption of multiple ciphertexts with 6 cores, which is observed to be the most time-consuming part in Pivot. 
The secure computations using SPDZ are not parallelized because the current SPDZ cannot express parallelism effectively and flexibly.
These partially parallelized versions are denoted as Pivot-Basic-PP and Pivot-Enhanced-PP, respectively.
{The comparison with the baselines is reported in Section \ref{subsubsec:comparison}.}
Table \ref{table:parameters} describes the ranges and default settings of the evaluated parameters.

\begin{figure*}[t]
\begin{subfigure}[b]{.49\columnwidth}
  \centering
  \includegraphics[width=0.95\columnwidth]{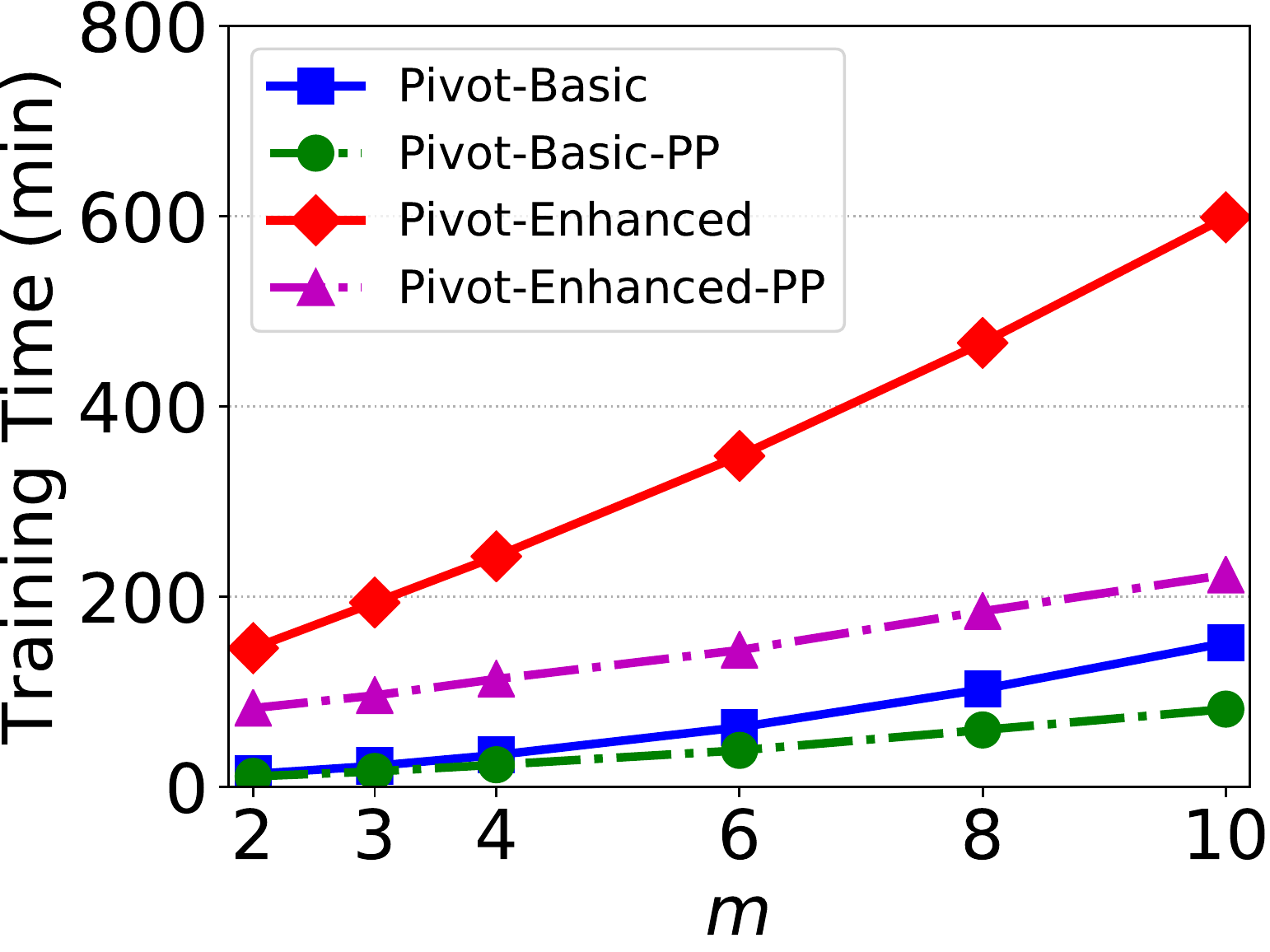}  
  \caption{{Training time vs. $m$}}
  \label{subfig:training-time-client-num}
\end{subfigure}
~
\begin{subfigure}[b]{.49\columnwidth}
  \centering
  \includegraphics[width=0.95\columnwidth]{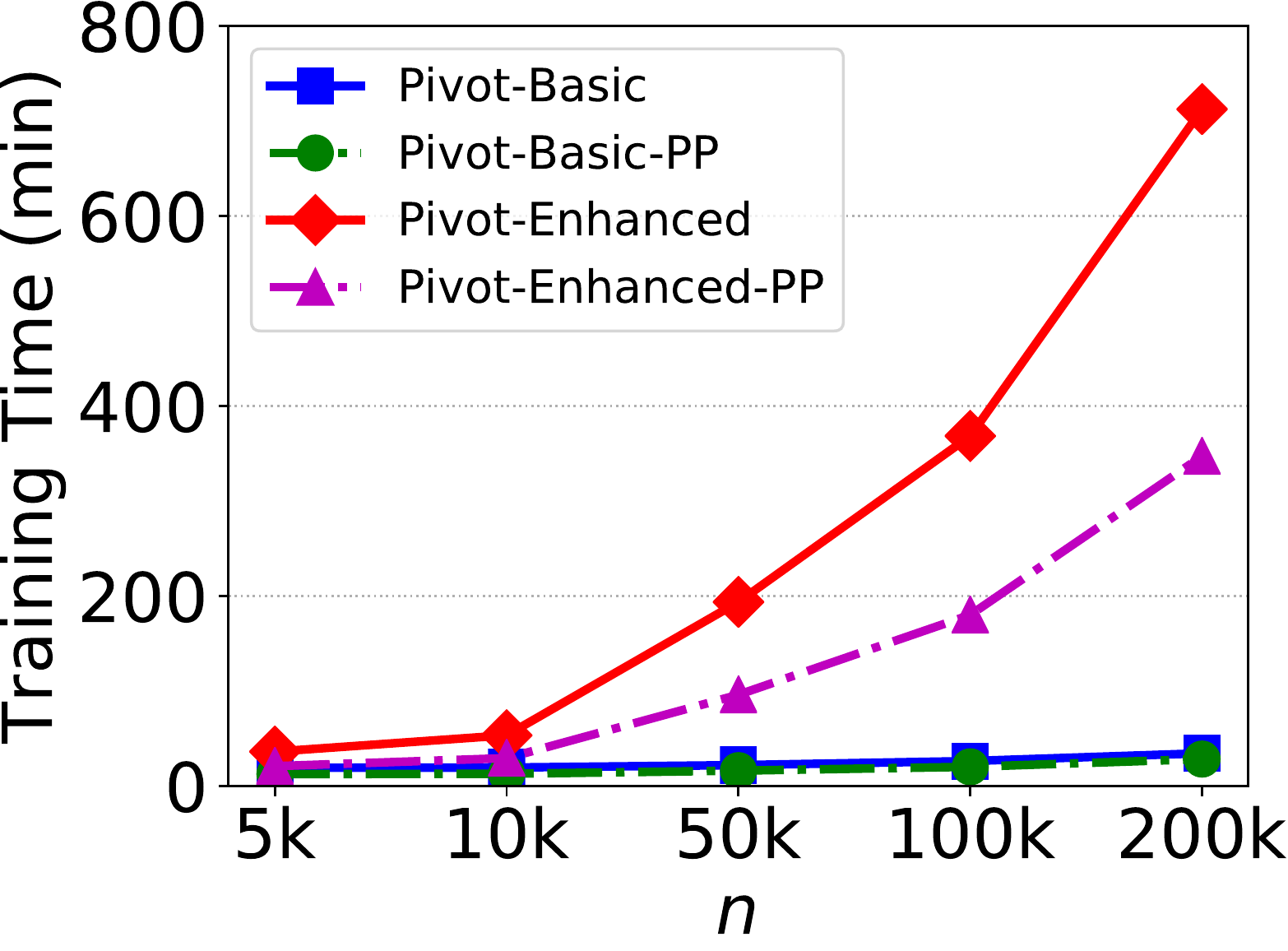} 
  \caption{Training time vs. $n$}
  \label{subfig:training-time-sample-num}
\end{subfigure}
~
\begin{subfigure}[b]{.49\columnwidth}
  \centering
  \includegraphics[width=0.95\columnwidth]{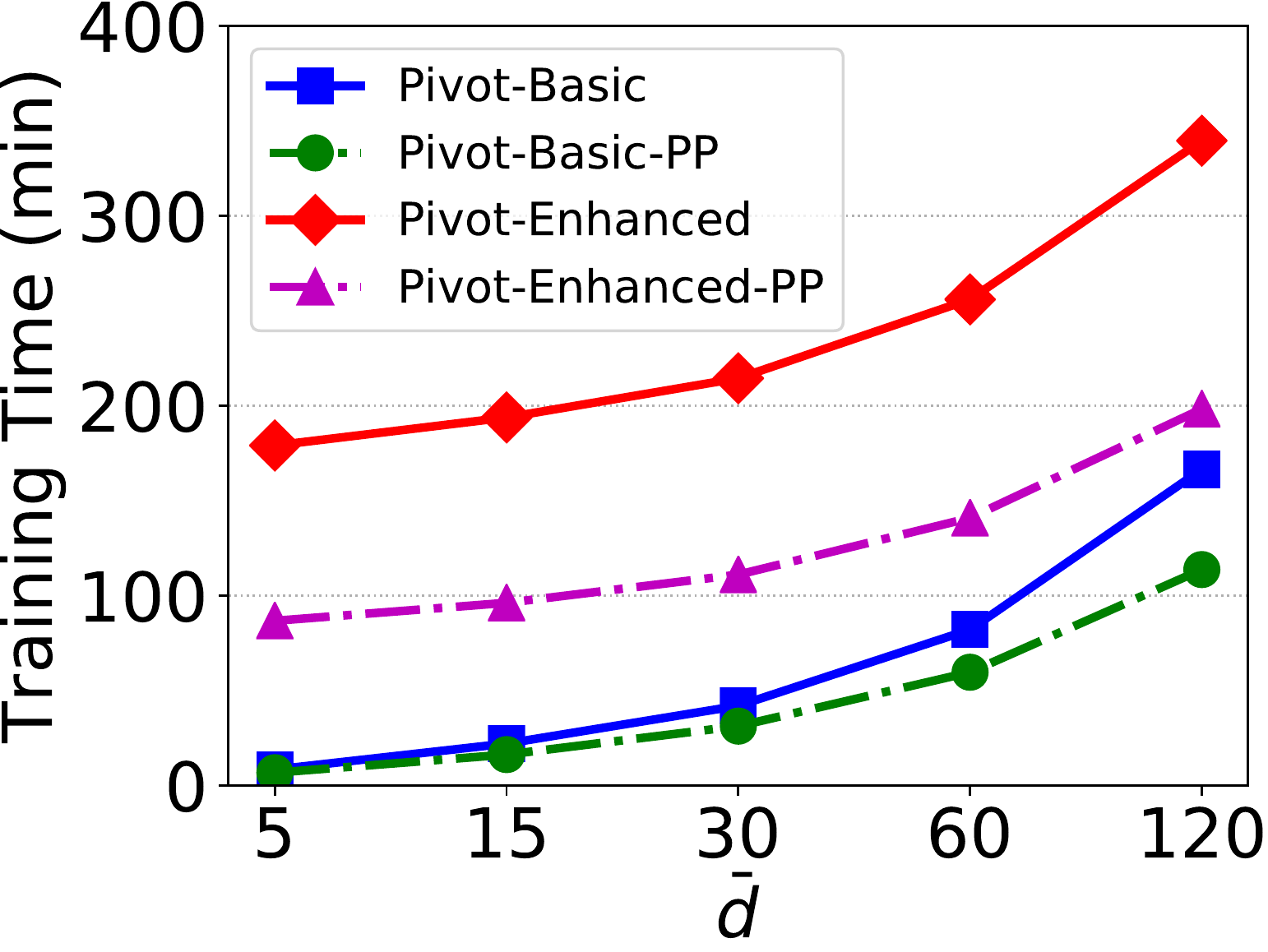}  
  \caption{{Training time vs. $\bar{d}$}}
  \label{subfig:training-time-feature-num}
\end{subfigure}
~
\begin{subfigure}[b]{.49\columnwidth}
  \centering
  \includegraphics[width=0.95\columnwidth]{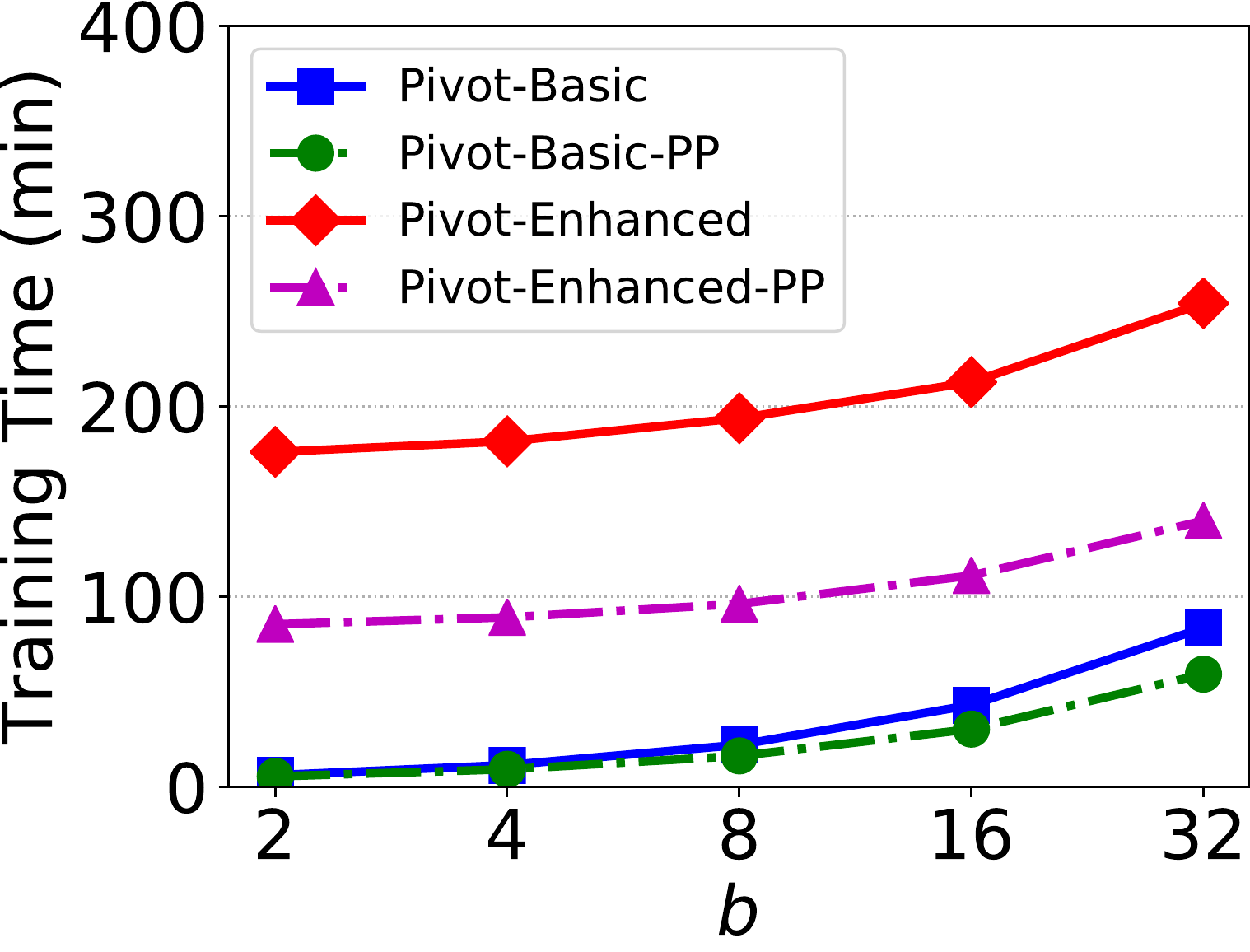}
  \caption{Training time vs. ${b}$}
  \label{subfig:training-time-split-num}
\end{subfigure}

\begin{subfigure}[b]{.49\columnwidth}
  \centering
  \includegraphics[width=0.95\columnwidth]{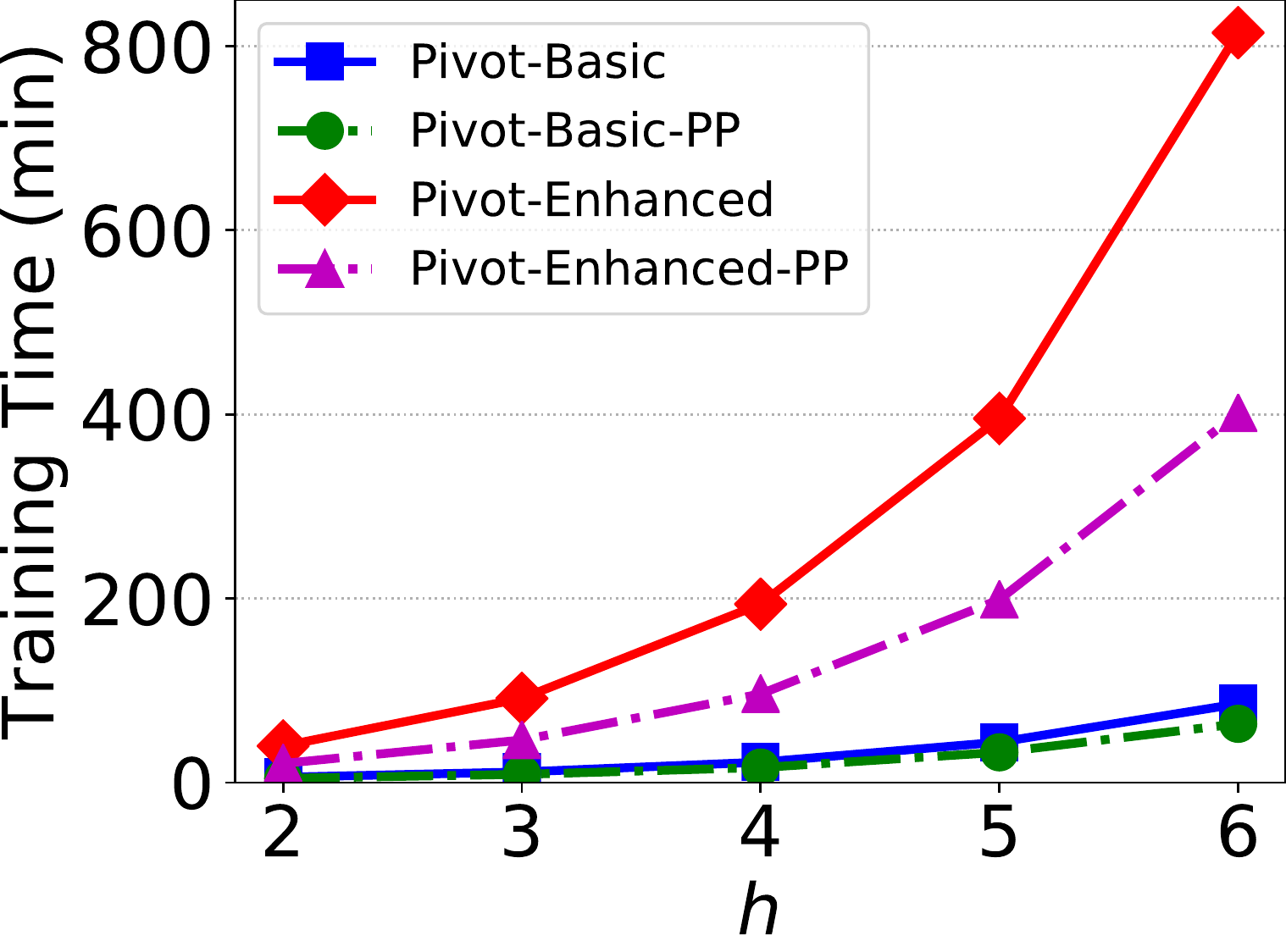}
  \caption{Training time vs. $h$}
  \label{subfig:training-time-tree-depth}
\end{subfigure}
~
\begin{subfigure}[b]{.49\columnwidth}
  \centering
  \includegraphics[width=0.93\columnwidth]{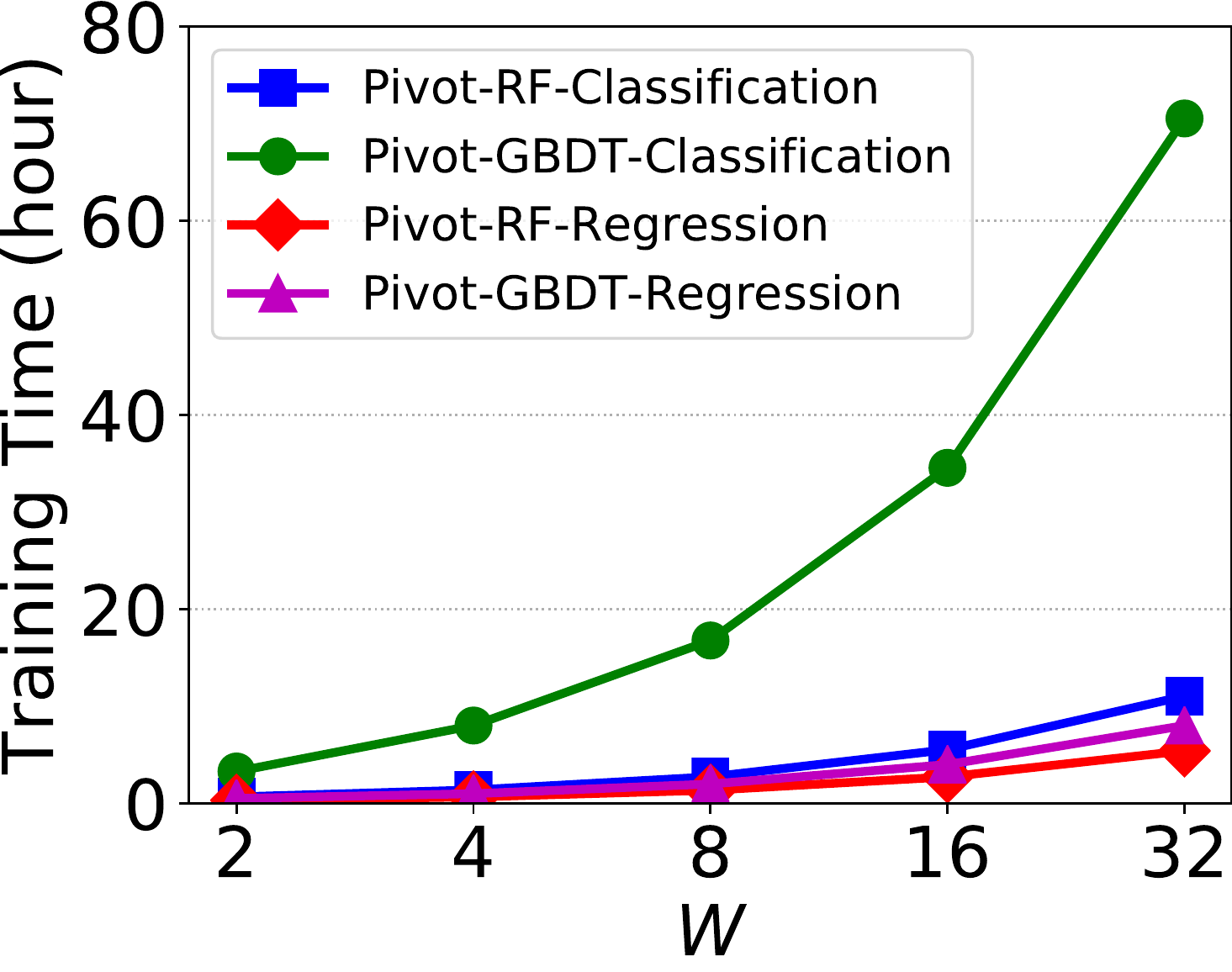} 
  \caption{{Training time vs. $W$}}
  \label{subfig:training-time-num-tree}
\end{subfigure}
~
\begin{subfigure}[b]{.49\columnwidth}
  \centering
  \includegraphics[width=0.93\columnwidth]{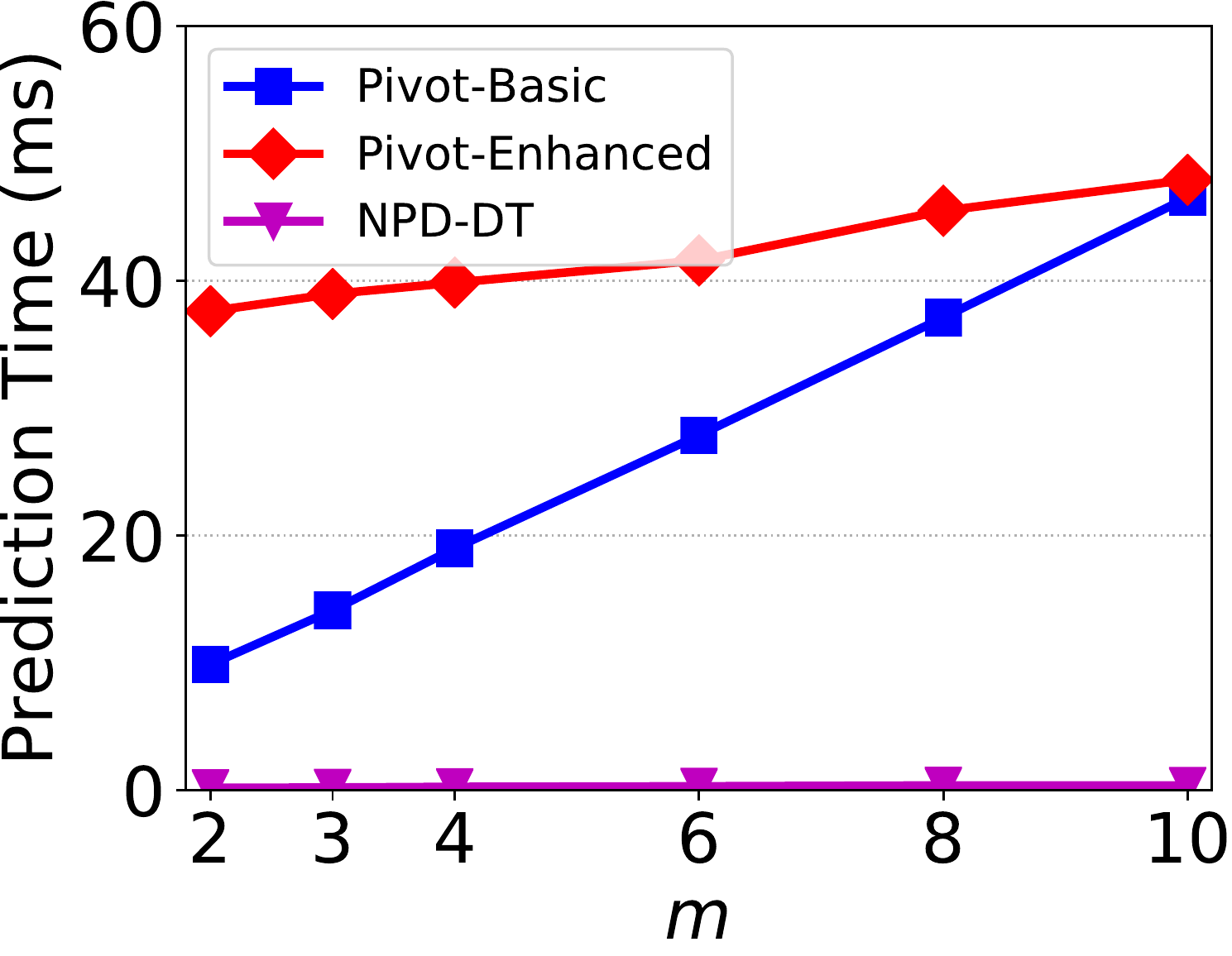}  
  \caption{{Prediction time vs. $m$}}
  \label{subfig:prediction-time-client-num}
\end{subfigure}
~
\begin{subfigure}[b]{.49\columnwidth}
  \centering
  \includegraphics[width=0.95\columnwidth]{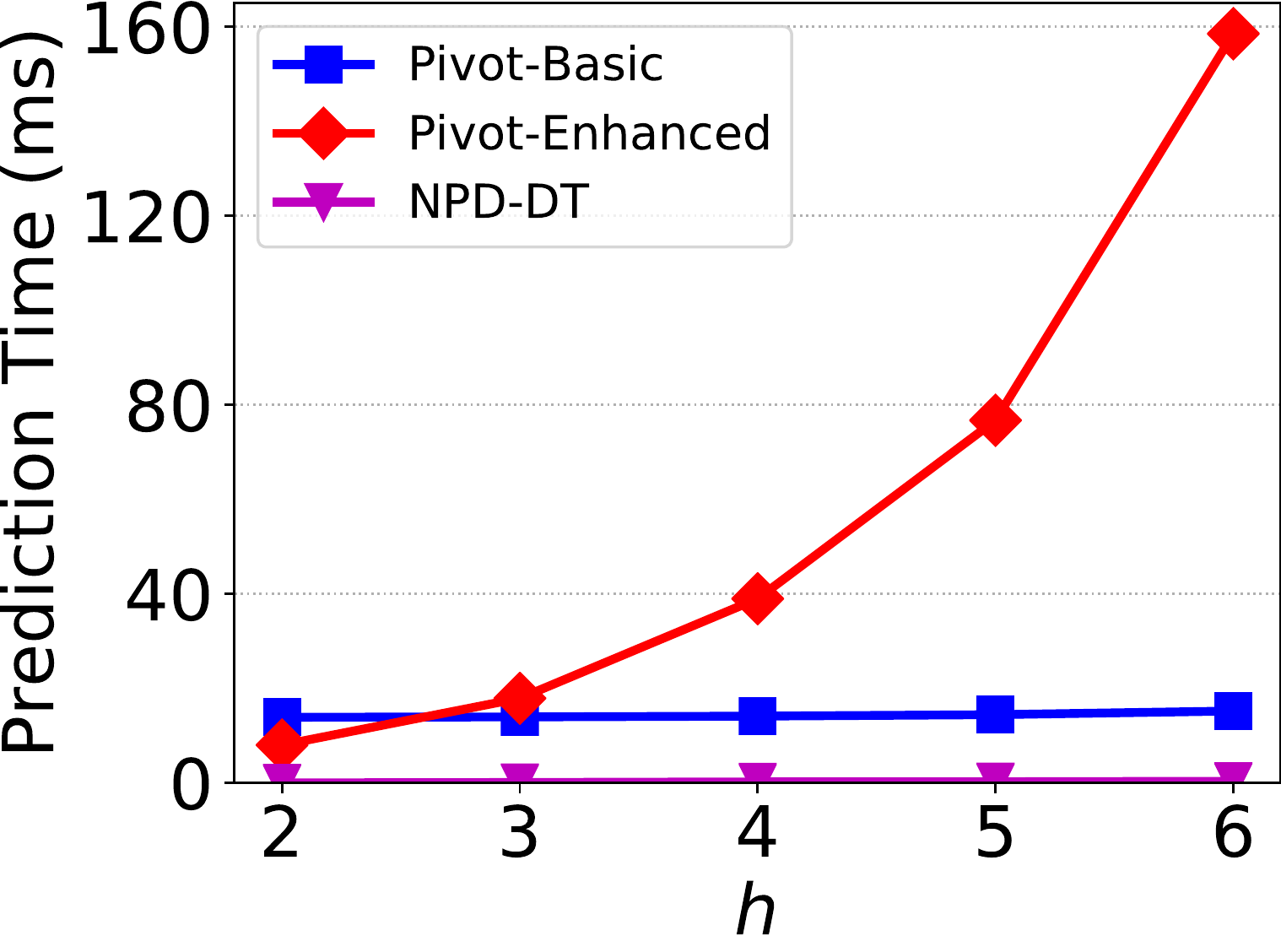} 
  \caption{Prediction time vs. $h$}
  \label{subfig:prediction-time-tree-depth}
\end{subfigure}

\caption{{Effect of parameters in decision tree models}}
\label{fig:effect-of-parameters}
\vspace{2mm}
\end{figure*}

\subsubsection{Evaluation on Training Efficiency}\label{subsubsec:evaluation-training}

\noindent
\textbf{Varying $m$.} 
Figure \ref{subfig:training-time-client-num} shows the performance for varying $m$. The training time of all algorithms increases as $m$ increases because the threshold decryptions and secure computations need more communication rounds.
Pivot-Basic always performs better than Pivot-Enhanced since Pivot-Enhanced has two additional computations in the model update step, where the $O(n)$ ciphertexts multiplications dominate the cost.
Besides, we can see that Pivot-Enhanced-PP that parallelizes only the threshold decryptions could reduce the total training time by up to 2.7 times. 

\vspace{1mm}
\noindent
\textbf{Varying $n$.} Figure \ref{subfig:training-time-sample-num} shows the performance for varying $n$. The relative comparison of the Pivot-Basic and Pivot-Enhanced is similar to Figure \ref{subfig:training-time-client-num}, except that the training time of Pivot-Basic increases slightly when $n$ goes up. The reason is that, in Pivot-Basic, the cost of encrypted statistics computation (proportional to $O(n)$) is only a small part of the total training time; the time-consuming parts are the MPC conversion that requires $O(c{d}b)$ threshold decryptions.
The training time of Pivot-Enhanced scales linearly with $n$ because of the threshold decryptions for encrypted mask vector updating are proportional to $O(n)$. For example, when $n = 200K$, the training time of Pivot-Enhanced is about 12 hours while that of Pivot-Basic is only 35 minutes.

\vspace{1mm}
\noindent
{\textbf{Varying $\bar{d}, b$.} Figure \ref{subfig:training-time-feature-num}-\ref{subfig:training-time-split-num} show the performance for varying $\bar{d}$ and $b$, respectively. The trends of the four algorithms in these two experiments are similar, i.e., the training time all scales linearly with $\bar{d}$ or $b$ since the number of total splits is $O(db)$. In addition, the gap between Pivot-Basic and Pivot-Enhanced is stable as $\bar{d}$ or $b$ increases. This is because that $\bar{d}$ does not affect the additional costs in Pivot-Enhanced, and $b$ only has small impact via private split selection (i.e., $O(n b)$ ciphertext computations) which is negligible comparing to the encrypted mask vector updating computation.}

\vspace{1mm}
\noindent
\textbf{Varying $h$.} Figure \ref{subfig:training-time-tree-depth} shows the performance for varying $h$. Since the generated synthetic datasets are sampled uniformly, the trained models tend to construct a full binary tree, where the number of internal nodes is $2^h - 1$ given the maximum tree depth $h$. Therefore, the training time of all algorithms approximately double when $h$ increases by one. 

\vspace{1mm}
\noindent
\textbf{Varying $W$.} Figure \ref{subfig:training-time-num-tree} shows the performance for varying $W$ in ensemble methods.
{
RF classification is slightly slower than RF regression as the default $c$ is 4 in classification comparing to $2$ in regression. GBDT regression is slightly slower than RF regression, since 
additional computations are required by GBDT to protect intermediate training labels. Besides, the training time of GBDT classification is much longer than GBDT regression for two overheads: one is the one-vs-the-rest strategy, which means $W*c$ trees are trained; the other is the \textit{secure softmax} computation on $c$ encrypted predictions for every sample in the training dataset, which needs additional MPC conversions and secure computations.}

\subsubsection{Evaluation on Prediction Efficiency}\label{subsubsec:evaluation-prediction}

\vspace{1mm}
\noindent
\textbf{Varying $m$.} Figure \ref{subfig:prediction-time-client-num} compares the prediction time per sample for varying $m$. Results show that the prediction time of Pivot-Enhanced is higher than Pivot-Basic, because the cost of secure comparisons is higher than the homomorphic computations. Besides, the prediction time of Pivot-Basic increases faster than that of Pivot-Enhanced as $m$ increases. The reason is that the communication round for distributed prediction in Pivot-Basic scales linearly with $m$; while in Pivot-Enhanced, the number of secure comparisons remains the same, the increasing of $m$ only incurs slight overhead.

\vspace{1mm}
\noindent
\textbf{Varying $h$.} Figure \ref{subfig:prediction-time-tree-depth} compares the prediction time per sample for varying $h$. When $h=2$, Pivot-Enhanced takes less prediction time because the number of internal nodes (i.e. secure comparisons) is very small. Pivot-Basic outperforms Pivot-Enhanced when $h\geq 3$ and this advantage increases as $h$ increases for two reasons. Firstly, 
the number of internal nodes is proportional to $2^h - 1$. Secondly, as described in Figure \ref{subfig:prediction-time-client-num}, the number of clients dominates the prediction time of Pivot-Basic; although the size of the prediction vector also scales to $h$, its effect is insignificant since the size is still very small, leading to stable performance.

\subsubsection{Comparison with Baseline Solution}\label{subsubsec:comparison}
{We compare the Pivot protocols with the baselines SPDZ-DT and NPD-DT. For NPD-DT, we report the training time for varying $m$ and $n$ in Figure \ref{subfig:compariosn-client-num}-\ref{subfig:comparison-sample-num}, and the prediction time per sample for varying $m$ and $h$ in Figure \ref{subfig:prediction-time-client-num}-\ref{subfig:prediction-time-tree-depth}. In all the evaluated NPD-DT experiments, the training time is less than 1 minute, and the prediction time is less than 1 ms. Nevertheless, the efficiency of NPD-DT is at the cost of data privacy.} For SPDZ-DT, since it is not parallelized, we adopt the non-parallelized versions 
We omit the comparison of prediction time, because the model prediction in SPDZ-DT is similar to that in Pivot-Enhanced. We compare with SPDZ-DT for varying $m$ and $n$. \par

\vspace{1mm}
\noindent
\textbf{Varying $m$.} Figure \ref{subfig:compariosn-client-num} shows the comparison for varying $m$. When $m = 2$, Pivot-Enhanced and SPDZ-DT achieve similar performance. However, the training time of SPDZ-DT increases much faster as $m$ increases because almost every secure computation in SPDZ-DT requires communication among all clients while most computations in Pivot protocols can be executed locally. {We can notice that Pivot-Basic and Pivot-Enhanced can achieve up to about 19.8x and 4.5x speedup over SPDZ-DT, respectively.} 

\vspace{1mm}
\noindent
\textbf{Varying $n$.} Figure \ref{subfig:comparison-sample-num} shows the comparison for varying $n$. Both Pivot-Enhanced and SPDZ-DT scale linearly to $n$ and SPDZ-DT increases more quickly than Pivot-Enhanced. When $n$ is small (e.g., $n = 5K$), the three algorithms achieve almost the same performance. While when $n = 200K$, Pivot-Basic and Pivot-Enhanced are able to achieve about 37.5x and 1.8x speedup over SPDZ-DT. 

%% file: tex-files/sec-further-extension.tex
\section{Further Protections}\label{sec:further-extensions}

This section extends Pivot to account for malicious adversaries (in Section~\ref{subsec:malicious-model}), and to incorporate {differential privacy} for enhanced protection (in  Section~\ref{subsec:differential-privacy}).

\subsection{Extension to Malicious Model}\label{subsec:malicious-model}

We demonstrate how to extend Pivot to account for malicious adversaries. Recall that we assumed a semi-honest adversary in Pivot, which means the clients do the executions correctly. In the malicious model, an adversary may deviate from the specified protocol to infer the private data. 
For example, in Algorithm \ref{alg:mpc-conversion}, if $u_1$ only adds its own encrypted share $[r_1]$ to compute $[e]$ (line 4), then it can infer the private data $x$ after the threshold decryption. 
To prevent such malicious behaviors, we let each client prove that it executes the specified protocol on the correct data (i.e., the data a client promises to use) step by step. Once a client deviates from the protocol or uses incorrect data in any step, the other clients will detect it and abort the execution. 

For this purpose, we extend Pivot using zero-knowledge proofs (ZKP) \cite{CramerDN01, Damgard2010} and authenticated shares in SPDZ \cite{DamgardPSZ12, KellerOS16}, inspired by \cite{Zheng2019}. We first present some building blocks in Section \ref{subsubsec:malicious-building-blocks}, then we introduce the extension to the basic protocol and the enhanced protocol in Section \ref{subsubsec:malicious-basic} and Section \ref{subsubsec:malicious-enhanced}, respectively. 

\subsubsection{Building Blocks}\label{subsubsec:malicious-building-blocks}
\noindent
\textbf{Zero-knowledge proofs (ZKP).} We use ZKP \cite{CramerDN01, Boudot00, Damgard2010, Zheng2019} to ensure that each client performs the local computation correctly, even if up to $m-1$ clients collude maliciously. Generally, ZKP enables a prover to prove to a verifier that a certain statement is true, without conveying any secret information for the statement. We mainly use the following existing building blocks of $\Sigma$-protocol for ZKP.
\begin{itemize}[topsep=2pt,itemsep=2pt,parsep=0pt,partopsep=0pt,leftmargin=15pt]
    \item Proof of plaintext knowledge (POPK): it takes a ciphertext $[a]$ as input and proves that the prover knows the plaintext $a_*$ such that $a_* = \textbf{Dec}([a])$ \cite{CramerDN01}.
    \item Proof of plaintext-ciphertext multiplication (POPCM): it takes three ciphertexts $[a], [b], [c]$ as input and proves that the prover knows the plaintext $a_*$ such that $a_* = \textbf{Dec}([a])$ and $\textbf{Dec}([c]) = \textbf{Dec}([a]) \cdot \textbf{Dec}([b])$ \cite{CramerDN01}.
    \item Proof of homomorphic dot product (POHDP): it takes two ciphertext vectors $[\boldsymbol{a}], [\boldsymbol{b}]$ and a ciphertext $[c]$ as inputs and proves that the prover knows the plaintext vector $\boldsymbol{a}_*$ such that $\boldsymbol{a}_* = \textbf{Dec}([\boldsymbol{a}])$ and $\textbf{Dec}([c]) = \textbf{Dec}([\boldsymbol{a}]) \odot \textbf{Dec}([\boldsymbol{b}])$ \cite{Zheng2019}. 
\end{itemize}
Note that the interactive $\Sigma$-protocol with honest verifier can be transformed into efficient non-interactive zero-knowledge (with random oracle assumption) and full zero-knowledge using existing techniques \cite{FaustKMV12, GarayMY03, Zheng2019}.

\begin{figure}[t]
\begin{subfigure}[b]{.49\columnwidth}
  \centering
  \includegraphics[width=0.95\columnwidth]{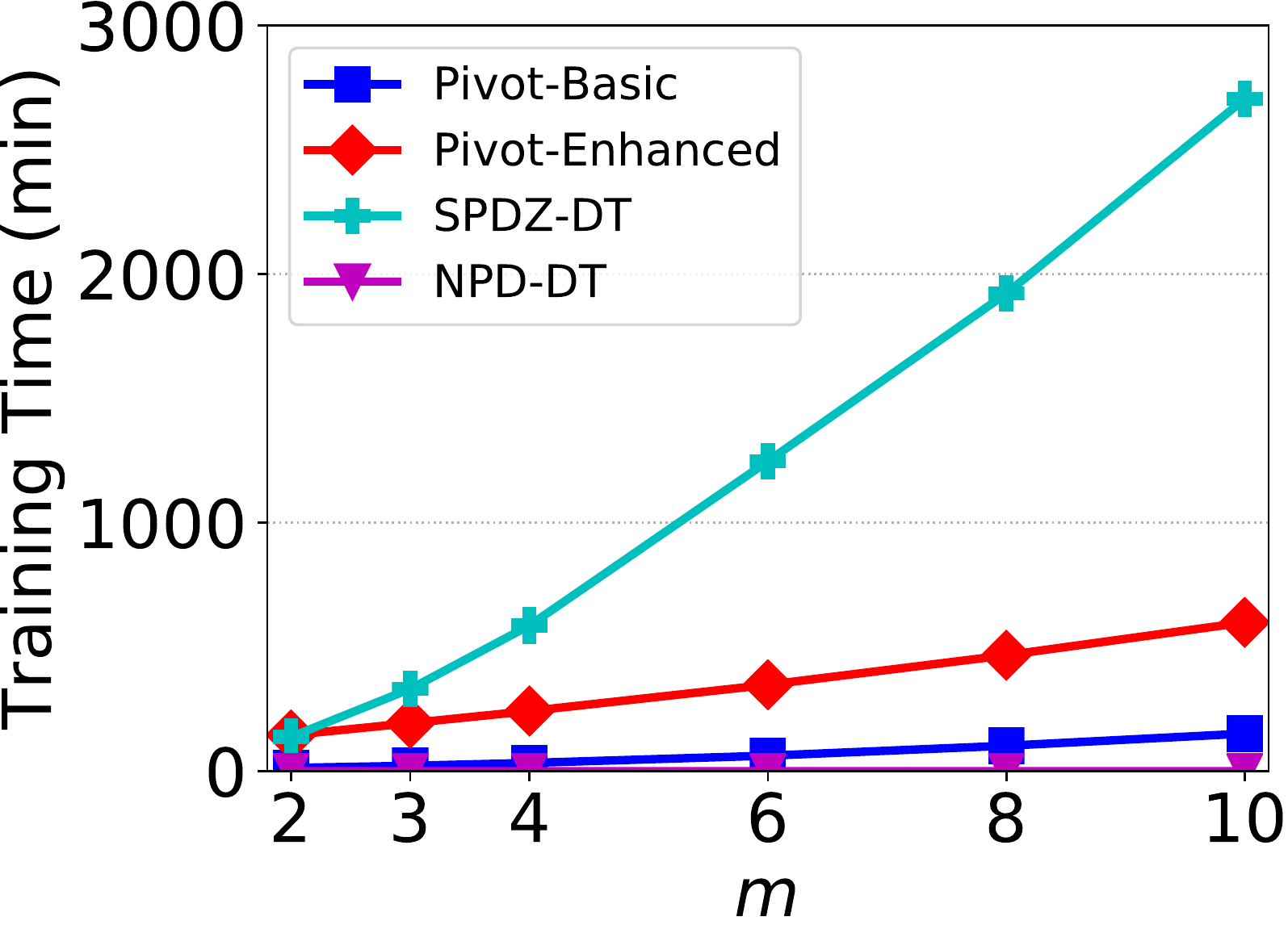}  
  \caption{{Training time vs. $m$}}
  \label{subfig:compariosn-client-num}
\end{subfigure}
~
\begin{subfigure}[b]{.49\columnwidth}
  \centering
  \includegraphics[width=0.95\columnwidth]{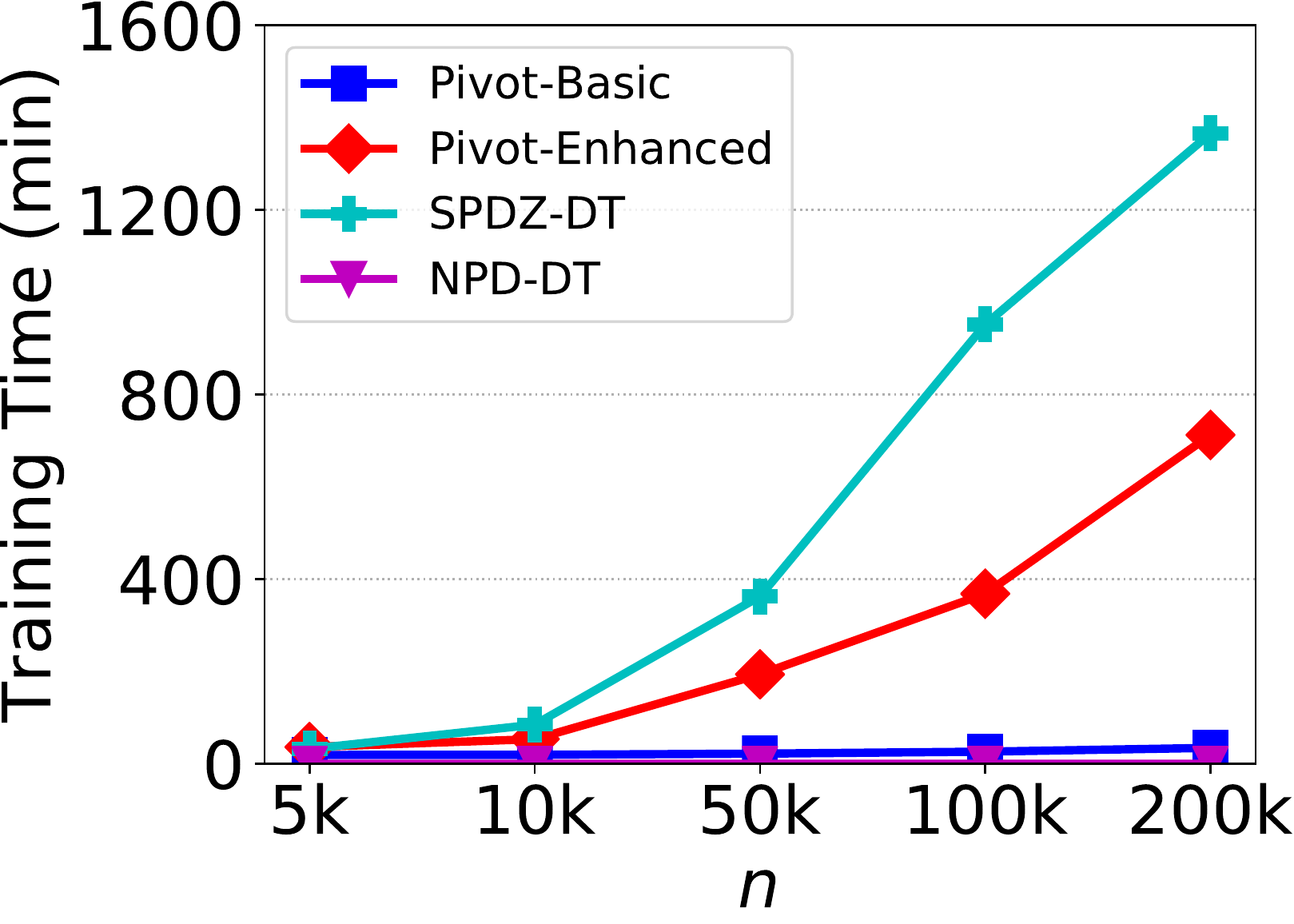}  
  \caption{{Training time vs. $n$}}
  \label{subfig:comparison-sample-num}
\end{subfigure}
\caption{{Comparison with baselines}}
\label{fig:comparison-with-spdz}
\vspace{2mm}
\end{figure}

\vspace{1mm}\noindent
\textbf{SPDZ authenticated shares.} SPDZ can ensure malicious security even up to $m-1$ clients may deviate arbitrarily from the protocol using the information-theoretic message authentication code (MAC) \cite{DamgardPSZ12, KellerOS16}.
The secure computation building blocks described in Section \ref{subsec:preliminaries:secure-multiparty-computation} are supported accordingly. In SPDZ, given a value $a \in \mathbb{Z}_q$, its authenticated secretly shared value is represented by $\langle a \rangle = ({\langle a \rangle}_1, \cdots, {\langle a \rangle}_m, {\langle \delta \rangle}_1, \cdots, {\langle \delta \rangle}_m, {\langle \Delta \rangle}_1, \cdots, {\langle \Delta \rangle}_m)$, such that client $i$ holds the random share ${\langle a \rangle}_i$, the random MAC share ${\langle \delta \rangle}_i$ and the fixed MAC key share ${\langle \Delta \rangle}_i$, and the MAC relation $\delta = a \cdot \Delta$ holds. 
The MAC-related shares ensure that no client can modify ${\langle a \rangle}_i$ without being detected.
When reconstructing a secretly shared value $\langle a \rangle$, every client $i \in \{1, \cdots, m\}$ first broadcast their shares ${\langle a \rangle}_i$ and compute $a = \sum_{i=1}^m {\langle a \rangle}_i$. 
To ensure that $a$ is correct, every client then checks the MAC by computing and opening ${\langle \delta \rangle}_i - a \cdot {\langle \Delta \rangle}_i$, then checking these shares sum up to zero.
If the MAC is incorrect, then the malicious behavior can be detected. 

\vspace{1mm}\noindent
\textbf{Modified MPC conversion.} Since Pivot applies a hybrid framework of TPHE and MPC, we need to modify Algorithm 1 as follows to make the MPC conversion process satisfy malicious security \cite{CramerDN01}. Specifically, we further let each client $i$: (i) broadcast $[r_i]$ together with POPK (line 3), ensuring that client $i$ knows $r_i$;
(ii) compute $[e]$ and call threshold decryption (line 4), ensuring that every client have computed the same $e$; 
and (iii) broadcast $[x_i]$ together with POPK (lines 6-8) for committing its own share. Then, the verifier can easily compute $[e-r_1]$ (if $i=1$) or $[-r_i]$ (if $i \neq 1$) using homomorphic properties (since both $[e]$ and $[r_i]$ are known to all), and check if it matches $[x_i]$ using a secure equality protocol under malicious model (e.g., \cite{KantarciogluK08}).

\subsubsection{Basic Protocol Extension}\label{subsubsec:malicious-basic}

We first discuss the extension of the classification tree training (Section \ref{subsec:classification-train}) and then the model prediction (Section \ref{subsec:prediction-trees}) of the basic protocol. The extension to the regression tree training follows the same way.

\vspace{1mm}\noindent
\textbf{Classification tree training.}
Before training, each client commits its local training data by encrypting and broadcasting it to other clients, which will be used for ZKP verification during the whole training process. 
The committed data includes the pre-computed split indicator vectors $\boldsymbol{v}_l$ and $\boldsymbol{v}_r$ for each local split, and the label indicator vector ${\boldsymbol{\beta}}_k$ of each class $k \in K$ that is only committed by the super client. 
Each client uses POPK to prove that it knows the plaintext of the committed data (e.g., $[\boldsymbol{v}_l]$, $[\boldsymbol{v}_r]$, $[\boldsymbol{\beta}_k]$).
Besides, the super client initializes an encrypted mask vector $[\boldsymbol{\alpha}]$ with $[1]$ and broadcasts it. 
It can be easily verified by threshold decryption since the initial $\boldsymbol{\alpha}$ is public.

\vspace{1mm}
\textit{Local computation.} 
In this step, the super client first computes and broadcasts a set of encrypted indicator vectors $[\BbbGammaVar] = \bigcup_{k\in K} \{[\boldsymbol{\gamma}_k]\}$ by element-wise homomorphic multiplication on ${\boldsymbol{\beta}}_k$ using $[\boldsymbol{\alpha}]$ for $k \in K$. 
Note that $[\boldsymbol{\beta}_k]$ has been committed by the super client and $[\boldsymbol{\alpha}]$ is also known to all, the super client can use POPCM to prove that she executes the homomorphic multiplication correctly. 
After that, each client computes $2c+2$ encrypted statistics for each local split, where the only operation is the homomorphic dot product computation, e.g., $[n_{l}] = \boldsymbol{v}_l \odot [\boldsymbol{z}]$. 
Therefore, each client can broadcast these encrypted statistics and prove that she performs these computations correctly using POHDP. 

\vspace{1mm}
\textit{MPC computation.} After utilizing the modified MPC conversion algorithm, the clients obtain random shares for the encrypted statistics.
To ensure that the random shares are not modified before combining with the MAC shares and computing with SPDZ, the clients also need to verify that these shares (along with the MAC shares) are valid and indeed match with the converted encrypted values \cite{Zheng2019}.
The rest of SPDZ computations are malicious secure as the authenticated secret sharing scheme is malicious secure, and the best split identifier $\langle i^* \rangle, \langle j^* \rangle, \langle s^* \rangle$ can be found and revealed to all clients. 

\vspace{1mm}
\textit{Model update.} In this step, client $i^*$ first selects the corresponding split indicator vectors $\boldsymbol{v}_l$ and $\boldsymbol{v}_r$ of the $s^*$-th split of the $j^*$-th feature, and computes $[\boldsymbol{\alpha}_l]$ and $[\boldsymbol{\alpha}_r]$ by element-wise homomorphic multiplication using $[\boldsymbol{\alpha}]$, which can be proved by POPCM. Note that the other clients can select the corresponding $[\boldsymbol{v}_l]$ and $[\boldsymbol{v}_r]$ for the verification given the best split identifier, as they have been committed beforehand. 

\vspace{1mm}
Similarly, the pruning condition check and leaf label computation can be proved. 
For example, if any pruning condition is satisfied, the super client can first compute and broadcast the encrypted number of samples for each class $k$ (say $[g_k]$) by summing up all elements in $[\boldsymbol{\gamma}_k]$, where $k  \in K$. Note that the verifier can execute the same computations (since $[\boldsymbol{\gamma}_k]$ is known to all) and use the secure equality protocol (e.g., \cite{KantarciogluK08}) to verify the correctness.
Then the clients convert them into shares, i.e., $\langle g_k \rangle$, and find the leaf label $\langle k \rangle$ that has the maximum $\langle g_k \rangle$ using the secure maximum operation (see Section \ref{subsec:classification-train}). The correctness can be ensured by the modified MPC conversion algorithm and the MAC-based SPDZ scheme. Finally, the leaf label can be revealed.

\vspace{1mm}
The above constructions guarantee that each client correctly follows the specified training protocol and uses the same data (i.e., split indicator vectors, and label indicator vectors, as committed before training) during the whole process. 
If the verification of any computation is incorrect, the execution will be aborted. 

\vspace{1mm}\noindent
\textbf{Model prediction.}
To ensure malicious security in the model prediction process, each client also needs to prove that she executes the specified computations using the correct data. 
Similar to the model training stage, each client can commit her data by encrypting and broadcasting the indicator of each sample's value comparing to the corresponding split threshold in the tree model. 
For example, the clients can commit their testing data for prediction along with the training data using the split indicator vectors; then a verifier can retrieve the other clients' committed indicators (for both left branch and right branch) given the sample index, client index, feature index and split index.

As described in Algorithm \ref{alg:private-decision-trees-prediction}, the clients execute the prediction process in a round-robin manner. At first, client $m$ initializes an encrypted prediction vector $[\boldsymbol{\eta}] = ([1], \cdots, [1])$ with size $t+1$ and broadcasts it to the other clients, where $t$ is the number of internal nodes. 
The clients can jointly decrypt $[\boldsymbol{\eta}]$ to check the correctness since the elements in this vector are known to all at the beginning. 
Then client $m$ updates $[\boldsymbol{\eta}]$ using her local sample indicators. For example, in Figure \ref{fig:tree-prediction}, there is one feature in the tree model that belongs to $u_3$, and the indicators are 0 (for the left branch) and 1 (for the right branch). 
For any tree node in the model, the client updates the corresponding leaf indexes in $[\boldsymbol{\eta}]$ using the indicators. 
For example, $u_3$ updates the first and fourth elements by homomorphic multiplication using 0 while the second and fifth elements by homomorphic multiplication using 1. 
For any update, the client broadcasts the updated $[\boldsymbol{\eta}]$ together with POPCM such that the other clients can verify the correctness. 
After $[\boldsymbol{\eta}]$ is updated by the last client (i.e., $u_1$ in Algorithm \ref{alg:private-decision-trees-prediction}), each client can homomorphicly aggregate $[\boldsymbol{\eta}]$ and call threshold decryption to check if the sum is 1, ensuring that there is only one valid prediction path.
Consequently, each client computes the homomorphic dot product operation between $[\boldsymbol{\eta}]$ and the leaf label vector $\boldsymbol{z}$ and calls threshold decryption to get the prediction output. 

\subsubsection{Enhanced Protocol Extension}\label{subsubsec:malicious-enhanced}

\vspace{2mm}\noindent
\textbf{Model training.} The commitment and the local computation step are exactly the same as those of the basic protocol. In the MPC computation step, instead of revealing $\langle s^* \rangle$, the clients compute a secretly shared indicator vector $\langle \boldsymbol{\lambda} \rangle = ({\langle \lambda_1 \rangle}, \cdots, {\langle \lambda_{n'} \rangle})$ using SPDZ, where $\lambda_t = 1$ when $t=s^*$ and $\lambda_t = 0$ otherwise. This step is malicious secure since SPDZ is malicious secure. Then, for each value in $\langle \boldsymbol{\lambda} \rangle$, the clients convert it into an encrypted value, by encrypting and broadcasting each share, and homomorphicly aggregating them together. The clients also need to verify that each encrypted value indeed match with the converted secretly shared value \cite{Zheng2019}. \par

In the model update step, client $i^*$ first computes a private split selection operation using $[\boldsymbol{\lambda}]$ and $\boldsymbol{V}^{n \times n'}$, where $\boldsymbol{V}$ is the split indicator matrix for all the splits of the $j^*$ feature and has been committed before training.  
The private split selection actually executes $n$ homomorphic dot product operations, which can be proved using POHDP.
After that, client $i^*$ executes an element-wise ciphertext multiplication between the selected encrypted split indicator vector $[\boldsymbol{v}]$ and the encrypted mask vector $[\boldsymbol{\alpha}]$ (see Section \ref{subsec:hiding-split-label}). The correctness can be ensured by the modified MPC conversion algorithm, the homomorphic multiplication together with POPCM, and the conversion from secretly shared value to ciphertext (as discussed above).

Similarly, after obtaining the secretly shared leaf label, the clients can jointly convert it into ciphertext, instead of revealing it. Therefore, the model training satisfies malicious security.

\vspace{1mm}\noindent
\textbf{Model prediction.} Recall that in the enhanced protocol, the clients first convert the tree model (with an encrypted split threshold on each internal node and encrypted leaf label on each leaf node) into secretly shared tree model, as well as convert their input feature values into secretly shared form. The conversion can be performed by the modified MPC conversion algorithm together with additional verification of the SPDZ authenticated shares (as discussed in Section \ref{subsubsec:malicious-building-blocks}). After that, the rest of the computations can be executed using malicious secure SPDZ, and the prediction output can be obtained.

\begin{algorithm}[t]
\DontPrintSemicolon
\small
\KwIn{${\mu}$: location parameter,
${b}$: scale parameter, 
}
\KwOut{$\langle X \rangle$: secretly shared value}
{
    $\langle U \rangle$ $\leftarrow$ sample a uniformly random secretly shared value within $(-\frac{1}{2}, \frac{1}{2})$ using SPDZ \\ 
    initialize secretly shared values $\langle U_s \rangle$ and $\langle U_a \rangle$  \\
    \If{$\langle U \rangle > \langle 0 \rangle$} {
        $\langle U_s \rangle = \langle 1 \rangle$, $\langle U_a \rangle = \langle U \rangle$
    }
    \ElseIf{$\langle U \rangle = \langle 0 \rangle$} {
        $\langle U_s \rangle = \langle 0 \rangle$, $\langle U_a \rangle = \langle 0 \rangle$
    }
    \Else {
        $\langle U_s \rangle = \langle -1 \rangle$, $\langle U_a \rangle = \langle -U \rangle$
    }
    $\langle X \rangle = \mu - b \cdot \langle U_s \rangle \cdot \ln (1 - 2 \cdot \langle U_a \rangle)$ $\slash\slash$ compute using SPDZ \\
    return $\langle X \rangle$
}
\caption{Randomly sample a secretly shared value from Laplace distribution}\label{alg:laplace-noise}
\end{algorithm}

\subsection{Incorporating Differential Privacy}\label{subsec:differential-privacy}

We can incorporate differential privacy (DP) \cite{DworkR14, HeMFS17, Shokri15, Chowdhury19, WangXYZHSS019, HayRMS10} to provide further protection, ensuring that the released model (even in the plaintext form) leaks limited information about individual's private data in the training dataset.
In a nutshell, a computation is differentially private if the probability of producing a given output does not depend very much on whether a particular sample is included in the input dataset \cite{DworkR14, Shokri15}.
Formally, for any two datasets $D$ and $D'$ differing in a single sample and any output $O$ of a function $f$,
\begin{align}\label{eq:differential-privacy}
\text{Pr}[f(D) \in O] \leq e^{\epsilon} \cdot \text{Pr}[f(D') \in O] 
\end{align}
The parameter $\epsilon$ is the differential privacy budget that controls the tradeoff between the accuracy of $f$ and how much information it discloses.

In our case, $f$ trains a CART tree model with multiple iterations where a tree node is built in each iteration. 
In the centralized DP (CDP) setting, a typical method for training a differentially private CART tree model is to make three queries satisfy DP in each iteration \cite{FriedmanS10, FletcherI19}: (i) pruning condition query (check if the number of samples $\Bar{n}$ on a tree node is less than a threshold); (ii) non-leaf query (determine the best split as a whole query); and (iii) leaf query (compute the leaf label).
Moreover, recall that in Pivot, no intermediate information is disclosed other than the tree model (i.e., each tree node) to be released.
As a result, the executions of Pivot essentially mimic the CDP setting in a way similar to \cite{Chowdhury19}. 
Therefore, we can incorporate the above CDP method into Pivot to make the training differentially private. 
We briefly introduce the classification tree case as follows (the extension to the regression tree is similar).
\par

We assign a DP budget $\epsilon$ to each query.
{First}, the clients can compute $[\Bar{n}]$ by homomorphicly aggregating $[\boldsymbol{\alpha}]$ and convert it to secretly shared $\langle \Bar{n} \rangle$. 
Before checking the condition, the clients jointly add a secretly shared random noise $\langle Lap( \Delta /\epsilon) \rangle$ to $\langle \Bar{n} \rangle$ according to the Laplace mechanism \cite{DworkR14}, where $\Delta$ is the sensitivity of the query, denoting the largest possible difference that one sample can have on the output of the query. Here $\Delta = 1$ since the count query can affect the output by maximum 1. 
Note that the random noise can be easily generated in an MPC way since the required primitives are all supported in SPDZ, such that no client knows the plaintext noise. 
Algorithm \ref{alg:laplace-noise} describes how to sample a secretly shared value from Laplace distribution using SPDZ. 
There are two steps: (i) uniformly samples a secretly shared value $\langle U \rangle$ within $(-\frac{1}{2},\frac{1}{2})$ (line 1), the primitive is also supported in SPDZ \cite{SPDZLibrary,DamgardPSZ12}; and (ii) computes the secretly shared value $\langle X \rangle = \mu - b \cdot \text{sgn}(\langle U \rangle) \ln(1-2 \cdot |\langle U \rangle|)$ (line 2-9), where $\mu$ and $b$ are the location parameter and scale parameter of the Laplace distribution. According to the inverse transform sampling \cite{Devroye86, vogel02, LaplaceDistribution}, the result $\langle X \rangle$ follows the Laplace distribution with parameters $\mu$ and $b$. In our case, $\mu = 0$ and $b = \frac{\Delta}{\epsilon}$. Consequently, the clients obtain the desired secretly shared random noise to be added on $\langle \Bar{n} \rangle$, and no one learns the plaintext noise. 

\begin{algorithm}[t]
\DontPrintSemicolon
\small
\KwIn{$\{ \langle \text{score}_1 \rangle, \cdots, \langle \text{score}_R \rangle \}$: secretly shared scores \newline
${\epsilon}$: differential privacy budget \newline
$\Delta$: score function sensitivity
}
\KwOut{$\langle \text{index} \rangle$: secretly shared index}
{
    \For{$r \in [1,R]$} {
        $\langle \text{prob}_r \rangle = \exp \Big( \frac{\epsilon \cdot \langle \text{score}_r \rangle}{2\Delta} \Big)$ $\slash\slash$ compute secretly shared probability for each score
    }
    $\langle \text{P} \rangle = \sum_{r=1}^R \langle \text{prob}_r \rangle$ \\
    $\langle F_0 \rangle = \langle 0 \rangle$ \\
    \For{$r \in [1,R]$} {
        $\langle \text{prob}_r' \rangle = \frac{\langle \text{prob}_r \rangle}{\langle P \rangle}$ $\slash\slash$ normalize the secretly shared probabilities such that their sum is $\langle 1 \rangle$
        $\langle F_r \rangle = \langle F_{r-1} \rangle + \langle \text{prob}_r' \rangle$ $\slash\slash$ cumulative probability for scores $\{1, \cdots, r\}$ \\
    }
    $(\langle 0 \rangle, \langle F_1 \rangle), (\langle F_1 \rangle, \langle F_2 \rangle), \cdots, (\langle F_{R-1} \rangle, \langle 1 \rangle)$ $\leftarrow$ arrange $R$ secretly shared sub-intervals within $(\langle 0 \rangle, \langle 1 \rangle)$ \\
    $\langle U \rangle$ $\leftarrow$ sample a uniformly random secretly shared value within $(0, 1)$ using SPDZ \\ 
    $\langle \text{index} \rangle = \langle -1 \rangle$ \\
    \For{$r \in [1, R]$}{
        \If{$\langle U \rangle > \langle F_{r-1} \rangle \wedge \langle U \rangle \leq \langle F_r \rangle$}{
            $\langle \text{index} \rangle = \langle r \rangle$
        }
        \Else{
            $\langle \text{index} \rangle = \langle \text{index} \rangle$
        }
    }
    return $\langle \text{index} \rangle$
}
\caption{Randomly select a secretly shared index using exponential mechanism}\label{alg:exponential-mechanism}
\end{algorithm}

{Second}, if the condition is not satisfied, the clients compute the best split as described in Section \ref{subsec:classification-train}, then the clients can jointly use the exponential mechanism \cite{DworkR14} to choose the best split where the sensitivity of the Gini impurity gain is $\Delta = 2$ \cite{FriedmanS10}. Algorithm \ref{alg:exponential-mechanism} describes the random selection using SPDZ based on the exponential mechanism, where the inputs are a number of $R$ secretly shared scores, the differential privacy budget, and the sensitivity of the score function. 
The clients first compute the secretly shared probabilities according to the exponential mechanism (line 1-2). Next, the clients normalize these probabilities such that the sum is $\langle 1 \rangle$; meanwhile, the clients compute the secretly shared cumulative probability for each index (line 3-7), which will be used for randomly sampling from discrete distribution (as what does in the exponential mechanism). After that, the clients arrange $R$ sub-intervals within $(0,1)$ according to the cumulative probabilities (line 7). Then they uniformly sample a secretly shared value $\langle U \rangle$ within $(0,1)$ (line 8), and find the sub-interval that $\langle U \rangle$ falls into (lines 9-14). The secretly shared index of the sub-interval follows the discrete distribution computed above \cite{DiscreteDistribution, Devroye86}, which satisfies the exponential mechanism. Importantly, all the computations are executed in an MPC way using SPDZ, such that the clients learn nothing. 
In our case, in particular, after computing the impurity gain for all possible splits, the clients obtain a set of secretly shared impurity gains, which can be viewed as scores of the splits. The clients can use Algorithm \ref{alg:exponential-mechanism} to decide the best split while satisfying DP.

{Third}, if the condition is satisfied, the clients compute the encrypted number of samples for each class $k \in K$, convert them into secretly shared values, and add Laplace noise $\langle Lap(\Delta/\epsilon) \rangle$ to each value using Algorithm \ref{alg:laplace-noise} before computing the leaf label, where $\Delta = 1$.
Since each class contains disjoint samples, the noise adding can be composed in parallel \cite{DworkR14}. 
Notice that the budgets of queries on different tree nodes on the same depth do not accumulate, as they are carried out on disjoint samples \cite{FriedmanS10}. Besides, each tree node consumes $2\epsilon$ budget since the pruning condition query is indispensable.
As a result, the training satisfies $\epsilon_B$-DP, where $\epsilon_B = 2(h+1)\epsilon$ and $h$ is the maximum tree depth \cite{FriedmanS10}. 
Similar to \cite{Chowdhury19}, the DP guarantee is under computational differential privacy \cite{MironovPRV09}, as the adversary in the MPC setting is assumed to be computationally bounded.

The integration of DP with the enhanced protocol is the same as the basic protocol since the two additional computations are independent of the DP operations.
Meanwhile, it has an additive protection effect when integrating DP with the enhanced protocol, because an adversary needs to reverse the concealed model first before obtaining the differentially private model. A noteworthy aspect is that, since both the Laplace noise generation (Algorithm~\ref{alg:laplace-noise}) and the random selection using exponential mechanism (Algorithm~\ref{alg:exponential-mechanism}) are computed using SPDZ, we can easily incorporate DP into the malicious model (see Section \ref{subsec:malicious-model}) by replacing the semi-honest SPDZ scheme with the authenticated SPDZ scheme.
\par

%% file: tex-files/sec-related-work.tex
\section{Related Work}\label{sec:related-work}

The works most related to ours are \cite{WangXSY06, VaidyaC05, VaidyaCKP08, VaidyaSFML14, ChengCorr19, LiuLL19, HuNYZ19} for privacy preserving vertical tree models. None of these solutions, however, achieve the same privacy guarantee as our solution. \cite{WangXSY06, HuNYZ19} assume that the super client's labels can be directly shared in plaintext with other clients, which obviously violates the privacy regulations. 
\cite{VaidyaC05, VaidyaCKP08, VaidyaSFML14, ChengCorr19, LiuLL19} allow that some intermediate information during the training or prediction process can be revealed in plaintext, which compromises the client's data privacy. 
For example, all these solutions assume that the available sample ids on a tree node is public, from which any adversarial client can easily infer that those samples belong to the same class (given a leaf node) or have similar feature values (given any node except the root node) with regards to the split feature on its parent node, with high probability. Also, \cite{ChengCorr19, LiuLL19} allow the split statistics for determining the best split to be revealed in plaintext to the super client, which discloses the client's data distribution. 

Meanwhile, although several general techniques may be applicable to our problem, they suffer from either privacy deficiency or inefficiency. Secure hardware (e.g., Intel SGX \cite{McKeenABRSSS13}) protects client's private data using secure enclaves \cite{OhrimenkoSFMNVC16, Zheng17}. However, it relies on the assumption of a trusted third party (e.g., Intel Corporation) and is vulnerable to side channel attacks \cite{XuCP15}, which is not acceptable to many organizations. 
While secure multiparty computation (MPC) \cite{Yao82b} could provide a strong privacy guarantee, training machine learning models using generic MPC frameworks is extremely inefficient \cite{Zheng2019}. Some other tailored MPC solutions (e.g., \cite{Mohassel17, NikolaenkoWIJBT13}) that propose to outsource client's data to non-colluding servers are unrealistic in practice since it is difficult to find those qualified servers convincing the clients. 
Our solution falls into the tailored MPC technique and does not rely on any external party, achieving accuracy comparable to the non-private solutions. 
\cite{Zheng2019} also uses a hybrid framework of TPHE and MPC, but it mainly focuses on linear models in horizontal FL, while our work addresses tree-based models in vertical FL and further considers the privacy leakages after releasing the model. 

Finally, there are a number of works on collaborative prediction \cite{CockDHKNPT19, BostPTG15, Gilad-BachrachD16, LiuJLA17} that consists of two parties, one is the server holds the private model and the other is the client holds private data. After prediction, nothing more than the predicted output is revealed to both parties. However, these solutions cannot be directly adopted in our problem. Since every client knows (a share of) the tree model and holds a subset of feature values in our setting. Although \cite{ChengCorr19, LiuLL19} consider the same vertical FL scenario as ours, their methods disclose the prediction path (see Section \ref{subsec:prediction-trees}) and thus leak client's data privacy along that path.
In contrast, our solution guarantees that no intermediate information other than the final prediction output is revealed.

%% file: tex-files/sec-conclusion.tex
\section{Conclusions}\label{sec:conclusion}

We have proposed Pivot, a privacy preserving solution with two protocols for vertical tree-based models.
With the basic protocol, Pivot guarantees that no intermediate information is disclosed during the execution.
With the enhanced protocol, Pivot further mitigates the possible privacy leakages occurring in the basic protocol.
To our best knowledge, this is the first work that provides strong privacy guarantees for vertical tree-based models. 
The experimental results demonstrate Pivot achieves accuracy comparable to non-private algorithms and is highly efficient.

%% file: tex-files/sec-acknowledgements.tex
\section*{ACKNOWLEDGEMENTS}\label{sec:acknowledgements}

We thank Xutao Sun for his early contribution to this work. This research is supported by Singapore Ministry of Education Academic Research Fund Tier 3 under MOEs official grant number MOE2017-T3-1-007.